\documentclass[usenatbib,dvipsnames]{aa}
\usepackage{subfig}
\usepackage[T1]{fontenc}
\usepackage{ae,aecompl}
\usepackage{graphicx} 
\usepackage{amsmath}  
\usepackage{amssymb}  
\usepackage{lscape}
\usepackage{rotating}
\usepackage{comment}
\usepackage{hyperref}
\hypersetup{
    colorlinks=true,
    linkcolor=blue,
    filecolor=blue,
    urlcolor=black,
    pdftitle={Overleaf Example},
    pdfpagemode=FullScreen,
    }
\usepackage{txfonts}
\usepackage{microtype}
\usepackage{xcolor} 
\usepackage{cancel,soul,ulem}

\newcommand{\chxmate}{CHEX--MATE}
\newcommand{\xmm}{XMM-Newton}
\newcommand{\chandra}{Chandra}
\newcommand{\meerkat}{MeerKAT}

\usepackage{lipsum}

\hypersetup{colorlinks,linkcolor={blue},citecolor={blue}}

\begin{document}
\title{The dynamics of the Anglerfish cluster}

\author{B.~Destefanis,\inst{1,2} 
M.~Balboni,\inst{1,3}
I.~Bartalucci,\inst{1,4}
M.~Annunziatella,\inst{ 1}
F.~Gastaldello, \inst{1}
S.~De Grandi, \inst{5}
S.~Ghizzardi, \inst{1}
C.~Grillo,\inst{1,2}
L.~Lovisari, \inst{1,6}
S.~Molendi, \inst{1}
M.~Rossetti \inst{1}
}

\institute{
INAF --
Istituto di Astrofisica Spaziale e Fisica Cosmica di Milano, via A. Corti 14, I-20133 Milano, Italy
\and
Dipartimento di Fisica, Università degli Studi di Milano, via Celoria 16, 20133 Milano, Italy
\and
DIFA – Università di Bologna, Via Gobetti 93/2, I-40129 Bologna, Italy
\and
Università degli studi di Roma ‘Tor Vergata’, Via della ricerca scientifica, 1, 00133 Roma, Italy
\and
INAF – Osservatorio Astronomico di Brera, via E. Bianchi 46, 23807 Merate (LC), Italy
\and
Center for Astrophysics $|$ Harvard $\&$ Smithsonian, 60 Garden Street, Cambridge, MA 02138, USA\\
\\\email{benedetta.destefanis@inaf.it} 
}

\abstract
{Merging galaxy clusters represent the ideal laboratory to test our understanding of the large scale structure formation history and the processes involved.
While many merging clusters have been identified, only a limited number have been studied in detail through multi-wavelength analysis and dynamical reconstruction, this type of analysis being crucial to account for projection degeneracies. 
This work investigates the merger dynamics of the massive and complex cluster MACS0600 using high spatial, $\sim 15$ arcsec, radio and X-ray datasets in combination with ancillary optical data. 
We analyze the cluster morphology and the thermodynamic properties of the intracluster medium (ICM) through \xmm{} and \chandra{} X-ray observations, and explore the non-thermal component via diffuse radio emission observed with \meerkat{}. We find a disturbed X-ray morphology with multiple substructures and a clear offset between the bulk of the radio emission and the X-ray peak. At the location of the X-ray peak, we detect a compact cool core surrounded by hotter gas and associated with a surface brightness discontinuity consistent with a cold front. The central region exhibits elevated temperatures and hosts most of the diffuse radio emission, suggesting merger-driven turbulence. Optical data further support a relative motion between the cool core and the main cluster along the line of sight. We conclude that MACS0600 is undergoing a merger in which a compact cool core has crossed the main, more massive cluster without being completely disrupted, while significantly perturbing the surrounding ICM. }

\titlerunning{Anglerfish paper}
\authorrunning{B. Destefanis}

\maketitle
\section{Introduction}\label{sec:intro}
Located at the nodes of the cosmic web, galaxy clusters grow hierarchically through episodic mergers of smaller-mass systems \citep[for a review see e.g.,][]{kravtsov2012}. These are highly energetic events \citep[e.g.,][]{Voit2005}, in which up to $10^{64}\hbox{ ergs}$ are dissipated through low mach number shocks ($\mathcal{M}\lesssim 3$) and turbulence \citep[e.g.,][]{vanWeeren2019} on timescales of a few Gyrs. Therefore, characterizing merger dynamics and their effects is crucial for understanding the mechanisms governing the formation and evolution of galaxy clusters. 
Specifically, merging clusters serve as unique astrophysical laboratories to probe the evolution history of galaxy clusters and the physics involved. 

Clusters are mainly composed of dark matter and the main barionic component consists of a hot and diffuse gas named Intracluster Medium (ICM). This plasma permeates the space between galaxies and it is typically studied through space-based X-ray observations, which probe its thermal bremsstrahlung continuum and line emission \citep{sarazin1988}. On the other hand, radio observations \citep{Feretti2012} trace the non-thermal synchroton emission of relativistic electrons, accelerated through turbulence and shocks, in the presence of a magnetic field \citep{Brunetti2014}. These shocks and turbulence can also be detected as signatures in the X-ray emission, for example, temperature jumps, surface-brightness inhomogeneities \citep{Markevitch2007}. Therefore, disturbed X-ray morphologies and diffuse radio emission are direct evidence for cluster mergers, and combining observations \citep[e.g.,][]{Rahaman2020} from these two wavebands can lead to a deeper understanding of the merger dynamics.

Detailed multi-wavelength studies are essential to break the degeneracies introduced by projection effects and to reconstruct the three-dimensional geometry and kinematics of these collisions. An example of this approach is the study of the complex system MACS J0717.5+3745 by \citet{adam2017_a}: by leveraging a multi-band analysis that combined X-ray data with the Sunyaev-Zeldovich effect \citep[SZ, ][]{Sunyaev1972}, they were able to constrain the merger velocity along the line of sight, providing a more complete picture of the cluster's dynamical state. 
Another example of a combined analysis that helps elucidate the dynamical state of a cluster is \citet{Bartalucci2024}, which combines X-ray and optical observations to characterize the spatially resolved thermodynamical properties of the gas, as well as the spatial and velocity distribution of cluster galaxies. 
A growing number of recent studies have adopted similar multi-wavelength approaches to break projection effects and better constrain the three-dimensional structure and dynamics of galaxy clusters \citep[e.g.,][]{Golovich2016, Golovich2019, Girardi2005, Ruppin2020, Rahaman2020, Pandge2019}. 
Such comprehensive analyses are therefore essential to quantify the energy budget of the merger and for understanding how these energetic events impact the global scaling relations of the cluster population. 

\begin{figure*}[!ht]
\begin{center}
\begin{minipage}{0.5\textwidth}
    \includegraphics[width=\linewidth]{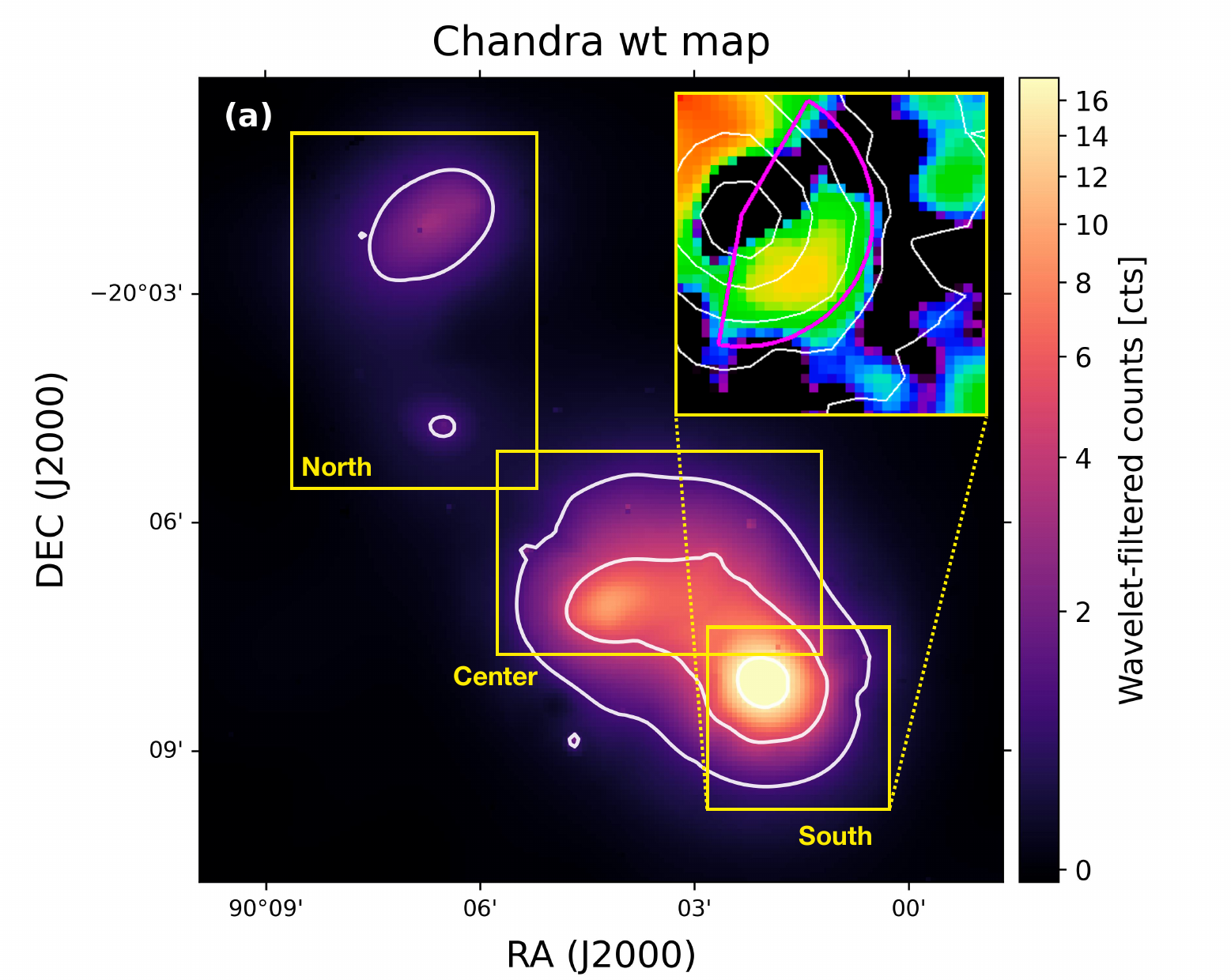}
\end{minipage}%
\begin{minipage}{0.5\textwidth}
    \includegraphics[width=\linewidth]{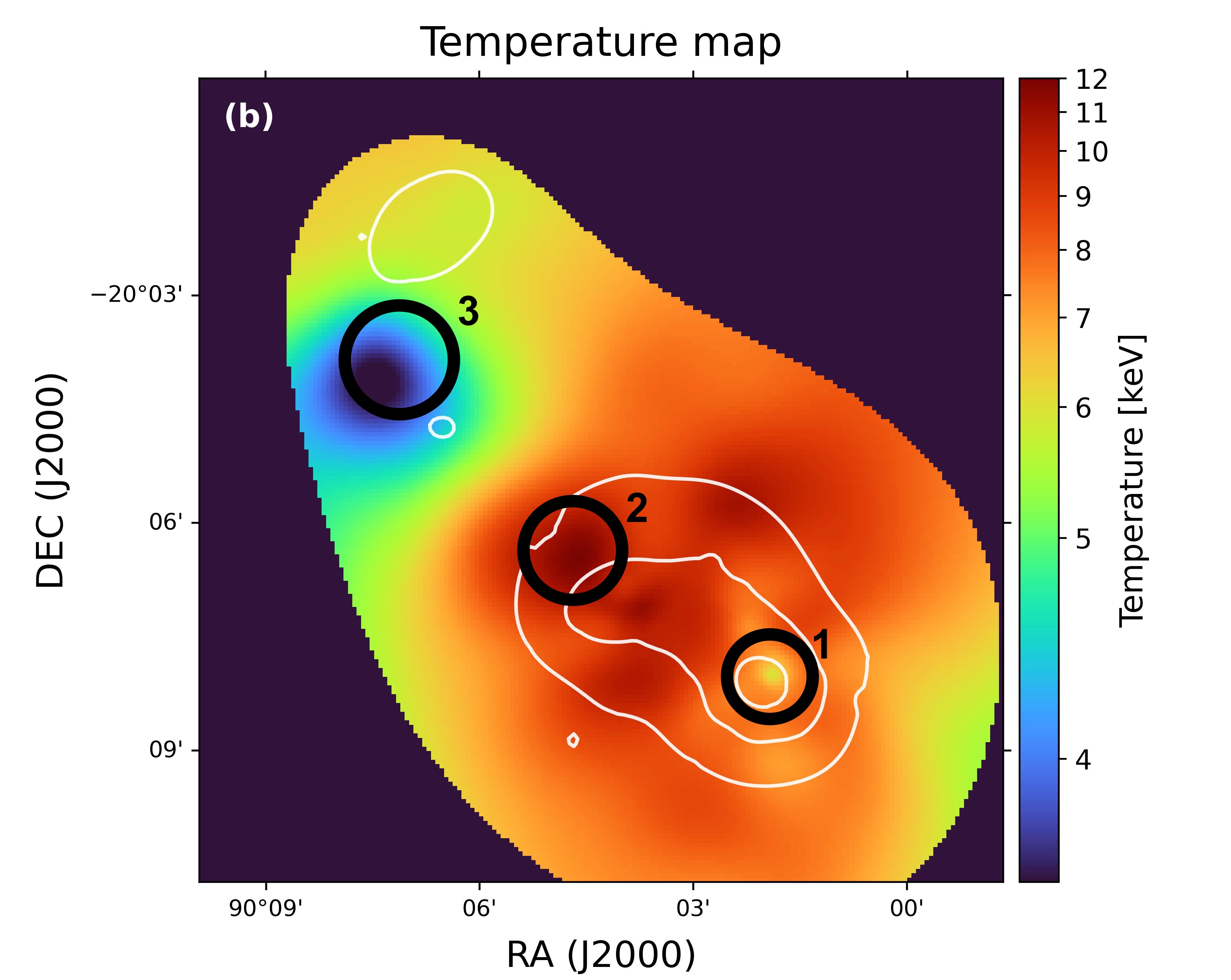}
\end{minipage}
\vspace{2pt}
\begin{minipage}{0.5\textwidth}
    \includegraphics[width=\linewidth]{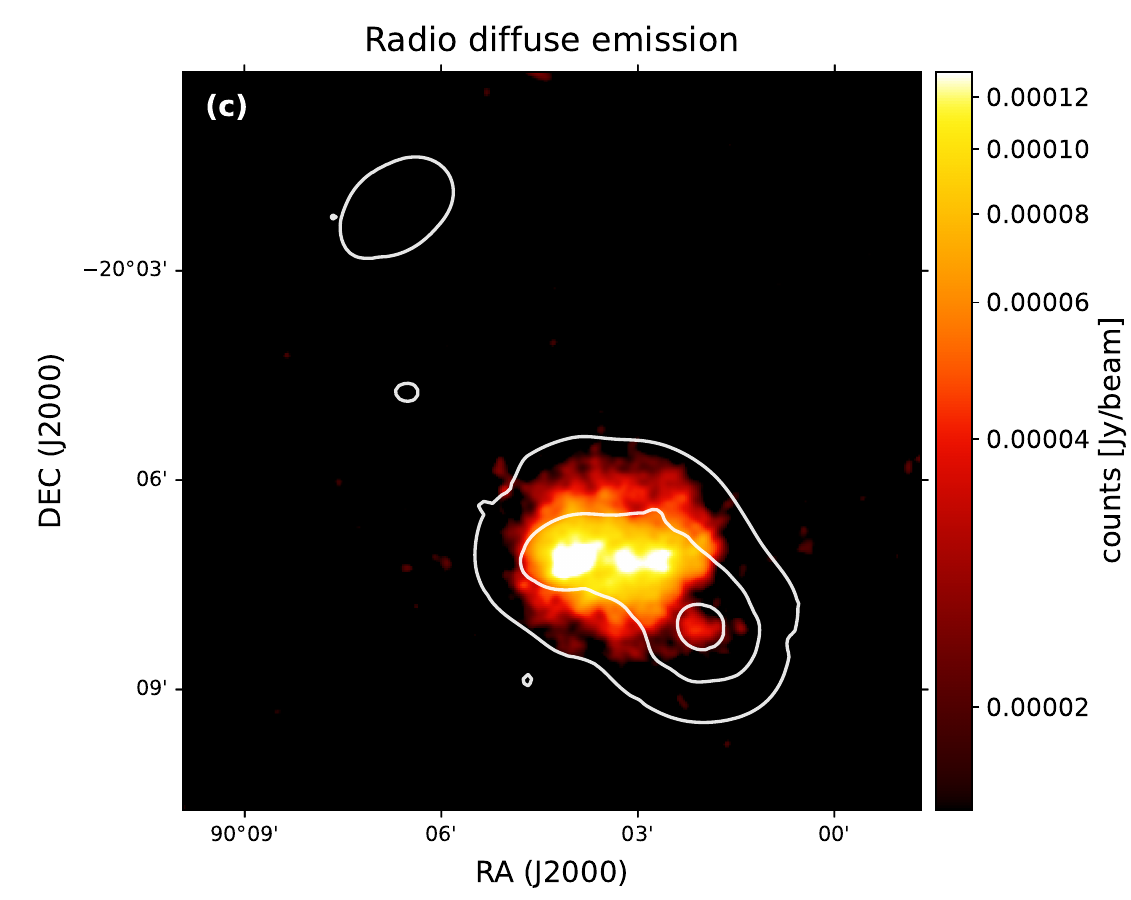}
\end{minipage}%
\begin{minipage}{0.5\textwidth}
    \includegraphics[width=\linewidth]{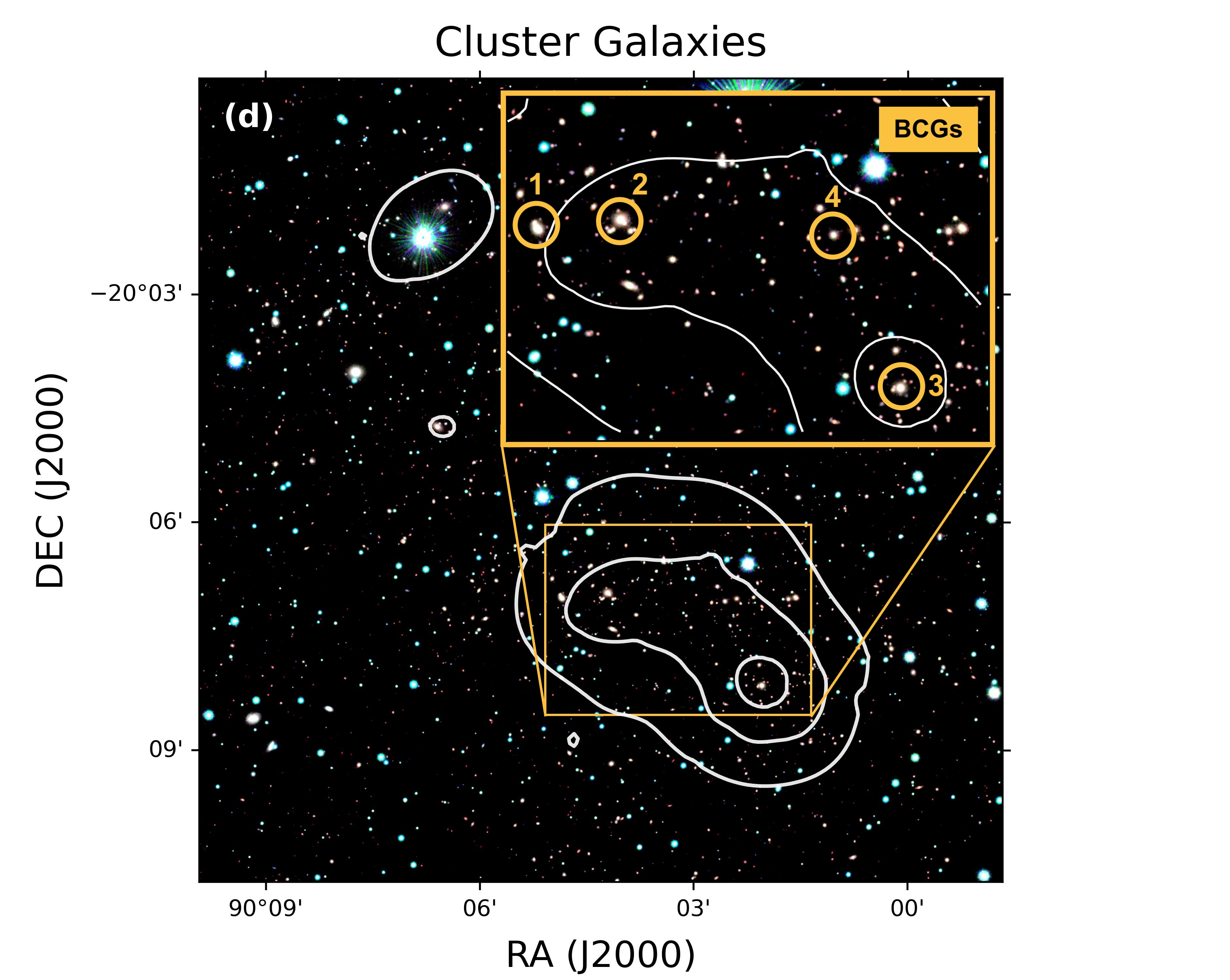}
\end{minipage}
\end{center}
\caption{
Global multiwavelength view of MACS J0600.1-2008. Top left: the X-ray SB map obtained from the \chandra{} dataset and filtered with the \texttt{wavelet} method to highlight the significant signal at $3\sigma$. In yellow are outlined three main regions on which our analysis focuses. In the inset, a zoom-in of the residual map is shown, with the average distance from the center at which the X-ray surface brightness discontinuity is observed marked in pink. Top right: the ICM temperature map obtained from the spectroscopic \xmm{} data, with the contours of the X-ray emission superimposed to provide spatial references. The regions, labeled from one to three, indicate those for which precise measurements of temperature are reported in this work (see Tab.~\ref{tab:temp}). Bottom left: diffuse radio emission from \meerkat{} (the image noise is $3.9\ \mu \hbox{Jy/beam}$ at a resolution of $9.2'' \times 8.9''$), with the contours of the X-ray emission superimposed. Bottom right: the RGB color image of the cluster galaxies, obtained using the three GCAV filters (blue: Y, green: J, red: Ks) and adopting the Lupton algorithm \citep{Lupton2004}. In the inset, the four BCGs, identified by \citet{Furtak2024} among the spectroscopically confirmed cluster member galaxies, are highlighted and labeled.
}
\label{fig:global_view}
\end{figure*}

At present, the highest-resolution spectroscopic and imaging observations of the ICM thermal emission are achievable with the \xmm{} and \chandra{} satellites, respectively. In this context, these two X-ray telescopes are perfectly complementary (e.g., the synergy between the two instruments for such cases has been shown in \citealt{adam2017_b}): \chandra{} has an unprecedented angular resolution \citep{Weisskopf2002}, allowing us to map the ICM morphology in great detail, while \xmm{} has a much higher throughput \citep{Jansen2001}, allowing us to derive spatial information of the temperature distribution. At radio wavelengths, recent years have seen significant advances in resolution and sensitivity. The most striking example is the results obtained using the \meerkat{} radio telescope \citep{Jonas2016}, which enables detailed studies of the substructures within diffuse radio emission in galaxy clusters \citep[e.g.,][]{Botteon2023}. However, deep studies of the dynamics of merging galaxy clusters remain limited, and systematic multi-wavelength comparisons are still poorly explored, highlighting the relevance of such analyses in the current state of the art. 
\vspace{0.3cm}
\newline
Motivated by this, the aim of this work is to investigate the dynamical state of a complex merging system through a detailed multi-wavelength analysis.
This study focuses on the merging cluster MACS J0600.1-2008 (\citealt{Ebeling2001}; also known as PSZ2 G225.93-19.99, \citealt{PlanckCollaboration2016}; dubbed Anglerfish Cluster by \citealt{Furtak2024}; MACS0600 hereafter). 
This cluster, at $z=0.432\pm0.007$ according to \citet{Furtak2024}, has been cataloged by \citet{Ebeling2001} among the most massive galaxy clusters in the Universe ($M_{500}^{SZ}=10.73^{+0.51}_{-0.54}\times 10^{14}\hbox{M}_{ \odot}$ from \citealt{PlanckCollaboration2016}). 
Here, $M_{500}$ denotes the enclosed mass corresponding to $R_{500}$, the radius within which the cluster mean density is $500$ times the critical density at the cluster redshift. 
MACS0600 center is located at right ascension $\mathrm{RA} = 06^{\mathrm h}00^{\mathrm m}11.3^{\mathrm s}$ and declination $\mathrm{Dec} = -20^\circ07^\prime14.5^{\prime\prime}$ (J2000) in the southern celestial hemisphere. 
The cluster belongs to the catalog of 118 clusters collected by the \chxmate{} program \citep{CHEXMATE2021}.
By combining four different X-ray morphological parameters, \citet{Campitiello2022} classified MACS0600 as the second (out of 116 analyzed) most disturbed CHEX-MATE object. 
Also \citet{Repp2018} indicate MACS0600 as a highly disturbed cluster among the MACS sample, inserting it into the most disturbed class of their optical morphology classification. Fig.~\ref{fig:global_view} presents a global view of the cluster, illustrating the different spatial distributions of the X-ray, radio, and optical emission. 
These properties make MACS0600 one of the most interesting objects to study for investigating merger dynamics and represents an exceptional laboratory for studying merger-driven processes in the ICM.

In addition, MACS0600 benefits from extensive multiwavelength coverage. In this paper, we analyze X-ray and radio observations and use optical data to support the interpretation of our results. 
In addition, the available X-ray data (\chandra{}) and radio observations (\meerkat{}) for this cluster are analyzed here for the first time.

The paper is organized as follows. In Section \ref{sec:sec2} we present the datasets used in this work, alongside with their cleaning and reduction processes. Section \ref{sec:sec3} describe our data analysis in the X-ray and radio bands. The results are then presented, discussed and interpreted in Section \ref{sec:sec4}. Finally, our conclusions are summarized in Section \ref{sec:conclusions}.

Throughout this work we assume a $\Lambda$CDM cosmological model, with Hubble parameter $H_{0}$ = 70 km s$^{-1}$ Mpc$^{-1}$, matter density $\Omega_{\rm m}$ = 0.3 and dark energy density $\Omega_{\Lambda}$ = 0.7.

\section{Observations and data preparation}\label{sec:sec2}
This study aims to constrain the dynamics of MACS0600. This is done by characterizing the X-ray morphology and the thermodynamic properties of the ICM, and by studying the non-thermal cluster component through radio observation.

The following sections describe the data used in this study and their preparation: the \xmm{} and \chandra{} X-ray observations in Section \ref{sec:dati-X}, the \meerkat{} radio data in Section \ref{sec:dati-meerkat}, and additional ancillary spectroscopy data in Section \ref{sec:dati-ottici}. 

\subsection{X-ray data}
\label{sec:dati-X}
As mentioned in Section \ref{sec:intro}, this work builds on the synergy between \xmm{} and \chandra{}, the first one being able to provide spatial information of the temperature distribution due to its high throughput, while the latter allowing to map the ICM surface-brightness (SB) with exquisite spatial resolution. 

\xmm{} observed MACS0600 on multiple occasions (observation IDs: 0650381401 and 0827050601), as did \chandra{} (observation IDs: 26075, 26101, 26102, 26103; doi:\href{https://doi.org/10.25574/cdc.679}{10.25574/cdc.679}), both offering a rather deep view of the cluster thermal emission. 
The two datasets were calibrated and reduced following the methodology described in detail in  \citet{Bartalucci2017} and \citet{Pratt2007}. In particular, the \xmm{} data were processed using the \xmm{} Science Analysis System (SAS\footnote{\href{https://www.cosmos.esa.int/web/xmm-newton/what-is-sas}{\xmm{} SAS}}) version 21.1.0 and the calibration files (CCF) as available to September 2025. From processed data we removed periods contaminated by flares and then applied a further cleaning filter to flag counts which are likely to be produced by the interaction of high energetic particles with the detector. After the cleaning stage, we extracted for each camera of \xmm{} (MOS1, MOS2, PN) the photon-count images in the [0.7-1.2] keV band \citep[i.e., the band that maximizes the signal-to-noise of the cluster thermal emission, ][]{Ettori2010} and then combined them to maximize the statistics. 
For the \chandra{} telescope the procedure is similar, using the Chandra Interactive Analysis of Observations \citep[CIAO, ][]{Fruscione2006} tools version 4.7 and the \chandra{}-ACIS calibration database version 4.6.5 of September 2025. 
After the reduction processes, the \xmm{} and the \chandra{} dataset effective exposure times are 50 ks and 120 ks, respectively.

When studying the X-ray emission from a galaxy cluster, the presence of bright point sources due to active galactic nuclei represents a significant contamination, both when dealing with X-ray images and spectra. Thus, point sources have been identified running on images in the [0.5-2.5] keV band the SAS tool \texttt{ewavelet} for \xmm{} dataset and the CIAO tool \texttt{wavdetect} for \chandra{}. The two lists of identified point sources were then merged together and checked by eye for false or missing detections. At the end of this process, the photons collected within the point source regions are removed from the analysis.

\subsection{Radio data}
\label{sec:dati-meerkat}
The \meerkat{} observations of MACS0600 were carried out to ensure 6 hours of on-source integration time 
\citep[PIDs: SCI-20230907-MB-02, ][]{Balboni2025}. These data had not been analyzed in detail before this study. In this work, we present the imaging and analysis of the L-band (900 - 1670 MHz) data.

The calibration of \meerkat{} observations followed here is described in detail in \citet{Balboni2025}, and it is divided in three main steps: (i) application of the Science Data Processor (SDP)\footnote{\href{https://skaafrica.atlassian.net/wiki/spaces/ESDKB/pages/338723406/SDP+pipelines+overview}{\meerkat{} SDP}.} Calibration Pipeline solutions; (ii) self-calibration of the data on the full field of view; (iii) extraction and self-calibration on the target. For the first step, the data were downloaded from the \meerkat{} archive using the "Default Calibration" option, which corrects the visibilities for bandpass, delay and gain calibration \citep{Hugo2021} and then converts the visibilities into \textsc{CASA} Measurement Set format \citep{McMullin2007, CASA2022}.  
Afterwards, the data were compressed and averaged in time and frequency by a factor of two. For the last two steps it was used \texttt{faceselfcal}. Where in the first part the data are self-calibrated on the full FoV, after an initial round of flagging on the Stokes V visibilities \citep{Botteon2024} using \textsc{AOFlagger} \citep{Offringa2010, Offringa2012} to remove radio frequency interference, and then a region is extracted around the target for the second round of self-calibration. 

The imaging was done with the w–stacking algorithm implemented in the \texttt{WSClean v3.6} software package \citep{Offringa2014}. To highlight the diffuse emission, we adopted a two-step approach. 
First, we created an image with a uv cut corresponding to a physical size of 100 kpc at the cluster redshift to obtain a model for the compact sources. Then we subtracted such a model from the visibility, obtaining a source-subtracted dataset. 
In the second step, we produced a low-resolution image using the \citet{Briggs1995} weighting scheme with \texttt{robust = –0.5} and applying a Gaussian uv taper in the visibility plane equivalent approximately to 50 kpc at the cluster redshift. 
The final image of the diffuse radio emission is presented in Fig.~\ref{fig:global_view}, panel (c).

\subsection{Optical and NIR data}     
\label{sec:dati-ottici}
Additional ancillary data for MACS0600, used in this work to aid the interpretation of our results, include the list of cluster members with spectroscopically confirmed redshifts (67 galaxies). 
Among these, \citet{Furtak2024} identifies four Brightest Cluster Galaxies (BCGs), labeled in panel (d) of Fig.~\ref{fig:global_view}. All the available spectroscopic redshifts, provided in the supplementary material of \citet{Furtak2024}, are gathered from multiple instruments, such as Gemini-N/GMOS and VLT/MUSE, as described therein.

Furthermore, the cluster galaxies image presented in panel (d) of Fig.~\ref{fig:global_view} is derived from the wide-field near-infrared (NIR) imaging taken with the VISTA infrared camera \citep[VIRCAM,][]{Dalton2006} in the framework of the Galaxy Clusters At VIRCAM survey (GCAV; Program ID: 198.A- 2008(E); PI: M. Nonino). The RGB color image is obtained using the three GCAV filters (blue: Y, green: J, red: Ks) and adopting the Lupton algorithm \citep{Lupton2004}. 
While the central region of the system has been extensively observed with both HST and JWST, the GCAV imaging dataset provides the only coverage of the full spatial extent of the cluster, enabling a complete view of its large-scale structure.

\section{Analysis}\label{sec:sec3}
Firstly, we performed a one-dimensional analysis of the cluster to characterize its global properties, using the \xmm{} dataset to exploit its higher statistics. 
In addition to the point sources, only for the one-dimensional analysis, we masked the emission coming from the northern sub-structure, so as to bring the cluster closer to the hypothesis of a relaxed and symmetric cluster. 
We extracted the SB radial profile centered on the X-ray peak (RA = $06^{\mathrm h}00^{\mathrm m}08.3^{\mathrm s}$, 
Dec = $-20^\circ08^\prime12.1^{\prime\prime}$, J2000) averaged over the entire azimuthal angle. We then obtained the density profile using the non-parametric approach of \citet{2006Croston} and derived the gas mass radial profile. With the gas mass in hand, we estimated $R_{500}=(1274 \pm 19)$ kpc and $M_{500}^{Y_X}=(9.62 \pm 0.43)\times10^{14}$ M$_{\odot}$ using the $Y_X$ proxy introduced by \citet{Kravtsov2006} as calibrated in \citet{arnaud2010}. 
The cluster mass $M^{Y_X}_{500}$ is very close to the mass obtained by the \citet{PlanckCollaboration2016} ($M^{SZ}_{500}=10.73^{+0.51}_{-0.54}\times 10^{14}\hbox{M}_{ \odot}$), thus confirming the system to be an exceptionally massive galaxy cluster. 

For the two-dimensional analysis of the \xmm{} and \chandra{} datasets we adopted the approach of \citet{Bourdin2004}, \citet{Bourdin2008}, \citet{Bourdin2013}, based on the construction of data-cubes from which one can easily extract two-dimensional maps and spectra. 
We extracted the X-ray SB maps and applied a wavelet filtering as detailed in \citet{Bourdin2013} to highlight the ICM emission at $3\sigma$ level (panel a in Fig.~\ref{fig:global_view}).

The implemented wavelet method was also used to obtain an ICM temperature map (panel b in Fig.~\ref{fig:global_view}), following the wavelet filtering algorithm described in \citet{Bourdin2004} and \citet{Bourdin2008}. It is worth noting that the spatial resolution of the temperature map shown depends on the statistics, since higher count rates provide better-constrained spectral fits of the ICM emission. For this reason, in the central parts where the cluster is brighter it is possible to appreciate the presence of structured temperature spatial distribution, while on the borders of the map the temperature structures are more diffuse. This wavelet-reconstructed temperature map was mainly used as a qualitative tracer of the thermal structure of the ICM and to identify regions of interest. All temperatures reported in this work were derived from dedicated local spectral fits, and not directly from the temperature map itself.

The different characteristics of \xmm{} and \chandra{} significantly impact these final products. While \chandra{} data have a much higher angular resolution, allowing the surface brightness map to contain finer details, \xmm{} data have higher photon statistics, resulting in a more accurate temperature map. From now on, any reference to the image analysis will imply the use of \chandra{} data, whereas spectral analysis will be based on \xmm{}. 

In panel (a) of Fig.~\ref{fig:global_view}, three regions are delineated: the southernmost one which includes X-ray peak, the central one and the northern one. In the detailed analysis of the cluster dynamics, we studied these three regions one by one. The considerations made on each are presented in the following paragraphs (\ref{sec:south}, \ref{sec:central}, \ref{sec:north}), while in the next section (Section \ref{sec:sec4}) we give an interpretation of the global scenario.

\subsection{South region}\label{sec:south}
The most prominent feature of this region is the the X-ray peak, around which the emission appears relatively smooth and nearly symmetric. This feature was tested through the analysis of the residuals.
Such a procedure was carried out by fitting a $\beta$-model \citep{cavaliere1976, Cavaliere1978} centered on the same X-ray peak considered in the one-dimensional analysis while masking the diffuse emission structures located north from the peak. Then the model was subtracted from the SB map to produce a residual map. 
The model used is of the form:
\begin{equation}
   I(r)=I_0 \biggl[  1+\bigg(\frac{r}{r_c}\bigg)^2  \biggl]^{-3\beta+1/2}\ \ + B,
\end{equation}
where the varying parameters are: $\beta$ is the slope, $r_c$ that is the core radius (in arcmin), $I_0$ that is the normalization parameter and $B$ that is the parameter that quantifies the background. 
We selected the azimuthal interval $212^{\circ} - 422^{\circ}$ to include only the southern region, where the emission is approximately circularly symmetric, and to exclude the diffuse substructures in the central and northern regions. The resulting parameter values are reported in Tab.~\ref{tab:fits}, obtained by fitting the profile extracted within a circular sector up to $3\hbox{ arcmin}$.
The inset present in panel (a) of Fig.~\ref{fig:global_view}, shows a zoom-in around the X-ray emission peak in the residual map. We note that the residuals exhibit a sharp transition from slightly positive to strongly negative values at approximately $1\hbox{ arcmin}$ from the peak (as highlighted by the pink sector). This may indicate a sudden emission decrease compatible with a gas density discontinuity. 
\begin{table}
    \caption{Values and errors for the $\beta$-model (left) and the three-dimensional broken power law projected along the line of sight (right) fit parameters}
    \label{tab:fits}
    \centering
    \renewcommand{\arraystretch}{1.3}
    \begin{tabular}{p{1.5cm} p{2cm} | p{1.5cm} p{2cm}}
        \hline
        \multicolumn{2}{c|}{\textbf{Beta Model}} & \multicolumn{2}{c}{\textbf{Broken Powerlaw Model}} \\
        \hline
        Parameter & Value & Parameter & Value \\
        \hline
        $\beta$ & $0.558\pm 0.014$ & $\alpha_1$ & $1.108\pm 0.010$ \\
        $r_c$ & $0.213\pm 0.015$ & $\alpha_2$ & $0.926\pm 0.033$ \\
        norm & $4.269\pm 0.027$ & $r_f$ & $0.926\pm 0.010$ \\
        bkg & $1.72\pm 0.06$ & norm & $3.010\pm 0.006$ \\
         &  & jump & $2.03\pm 0.11$ \\
         &  & bkg & $-1.6\pm 0.6$ \\
        \hline
    \end{tabular}
\end{table}

Another notable feature emerges from the temperature map: in correspondence of the X-ray brightness peak the gas appears to be cooler than the surrounding one. We extracted and fitted the spectrum in small regions inside the emission peak and in the rest of the cluster with \texttt{Xmap} \citep{Bourdin2004,Bourdin2008,Bourdin2013}. Three of the selected regions are outlined in black in panel (b) of Fig.~\ref{fig:global_view}, numbered one to three, and the temperature measures obtained in them are listed in Tab.~\ref{tab:temp}. 
Unfortunately, it was not possible to extract a radial temperature profile up to the outer region due to the low photon statistics even with \xmm{} data.
\begin{table}
    \caption{Temperature measures in three patches in the south, central, and north region}
    \centering
    \renewcommand{\arraystretch}{1.3}
    \begin{tabular}{c c c}
 \hline
 Regions&Temperature \\
 \hline
         1& $5.56_{-1.04}^{+1.44}\hbox{ keV}$ \\
         2& $10.18_{-2.50}^{+4.47} \hbox{ keV}$ \\
         3& $1.97_{-0.27}^{+0.72} \hbox{ keV}$ \\
 \hline
    \end{tabular}
    \renewcommand{\arraystretch}{1.0}
    \tablefoot{Here a spectrum was extracted and fitted using a standard spectral analysis, rather than temperatures reconstructed from the wavelet-based method. We used the abundance tables of \citet{anderson1989} during the fit procedure.}
     \label{tab:temp}
\end{table}

\begin{figure}
\centering
\resizebox{1\columnwidth}{!}{\includegraphics[]{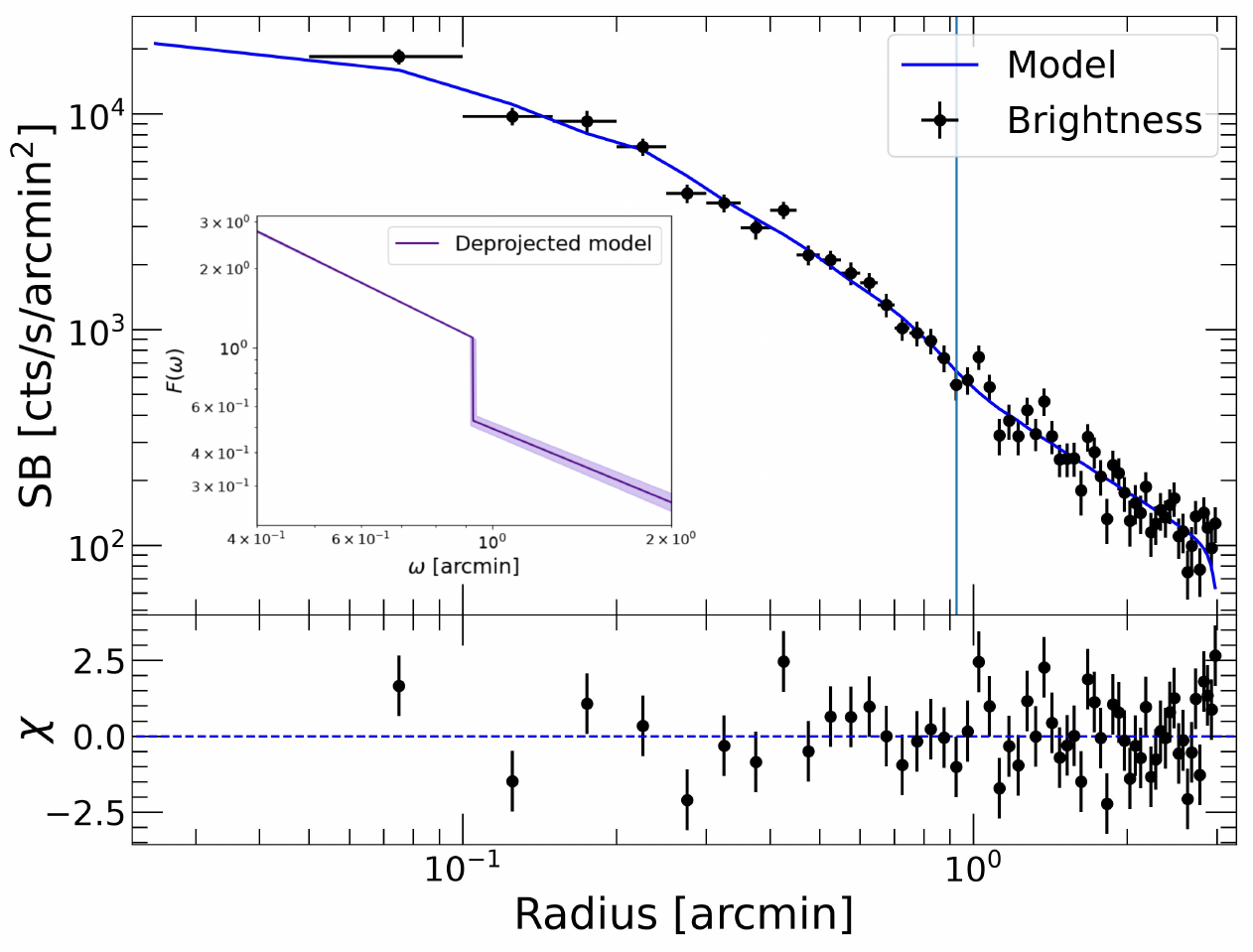}}
\caption{SB profile, obtained in a chosen sector in the southern region, with the fitted model on top (in blue). The vertical line marks the location of the discontinuity, at about $0.926 \hbox{ arcmin}$ from the X-ray peak. The inset shows the corresponding deprojected broken power law model (see Eq.~\ref{eq:bknPow_2}) using the best-fit parameters.}
\label{fig:jump}
\end{figure}

The temperature map and the residual map guided us in identifying the regions of discontinuity and in selecting the sector from which to extract the specific SB profile, using \texttt{pyproffit} \citep{Eckert2020}. Centering it on the X-ray peak, we derived the radial profile by averaging the SB value within concentric annular sectors up to $3\hbox{ arcmin}$, limiting the azimuthal interval $260^{\circ} - 420^{\circ}$. To fit such a profile we used a three-dimensional broken power law projected along the line of sight, a model often used to identify SB discontinuities \citep{Markevitch2007}. The model is
\begin{equation}
    I(r)=I_0\int F(\omega)^2\,\mathrm{d}l \ \ +\ B,
    \label{eq:BknPow}
\end{equation}
with $\omega^2=r^2+l^2$ and 
\begin{equation}
    F(\omega)=
\begin{cases}
\omega^{-\alpha_1}, & \omega < r_f \\
\dfrac{1}{C}\,\omega^{-\alpha_2}, & \omega \ge r_f
\end{cases}
\label{eq:bknPow_2}
\end{equation}
containing the following parameters: $\alpha_1$ and $\alpha_2$ which are the slopes of the two power laws; $r_f$ that is the value of the radius at which the discontinuity occurs (in arcmin); $C$ is called jump and it is the density compression factor, that quantifies the discontinuity; $I_0$ that is the normalization parameter and $B$ that is the parameter that quantifies the background. We fitted this model to the SB radial profile in the sector described above, and the resulting parameter values are reported in Tab.~\ref{tab:fits}. 
The fit clearly identifies a jump at $0.926 \hbox{ arcmin}$ from the X-ray peak, highlighted by the blue vertical line in Fig.~\ref{fig:jump}. 
This significant surface brightness jump strongly suggests the presence of a corresponding discontinuity in the gas density, consistent with a cold front. Thus, where the X-ray emission peaks, there is a cooler and compact gas core that has partially maintained its relaxed and symmetric state, but at the edges of which there is a significant SB discontinuity. The emerging scenario is compatible with a relatively cool core moving inside a hotter gas and generating a cold front ahead of it in the direction of motion. 

In the radio image it can be noted that, although most of the radio diffuse emission is concentrated in the central region of the cluster, a small radio trail follows the X-ray peak in the southern region. This is evident in panel (c) of Fig.~\ref{fig:global_view}, where this region of the radio low-resolution image is shown in correspondence with the X-ray emission contours. 
We verified that this is not a compact emission that mistakenly survived the subtraction of the compact sources. Indeed, a radio galaxy is present where the X-ray peak is, and coincides with the South BCG identified by \citet{Furtak2024}. However, the diffuse emission extends over much larger scales, and the subtraction of compact sources was performed correctly, since no other radiogalaxy contribution remained in the final image. 
Consequently, there is a clear tail of radio emission which, from the central region, seems to follow the X-ray peak, as if to trace a motion of this core starting from the center of the cluster toward the outside in the SW direction. Interestingly, this appears to be an increasingly common feature in merging systems, where diffuse radio emission follows displaced X-ray structures and traces the motion of infalling subclusters or cores \citep[e.g., in the Lyra complex, ][]{Botteon2019, Clavico2019}.

\begin{figure}
\centering
\resizebox{1\columnwidth}{!}{\includegraphics[]{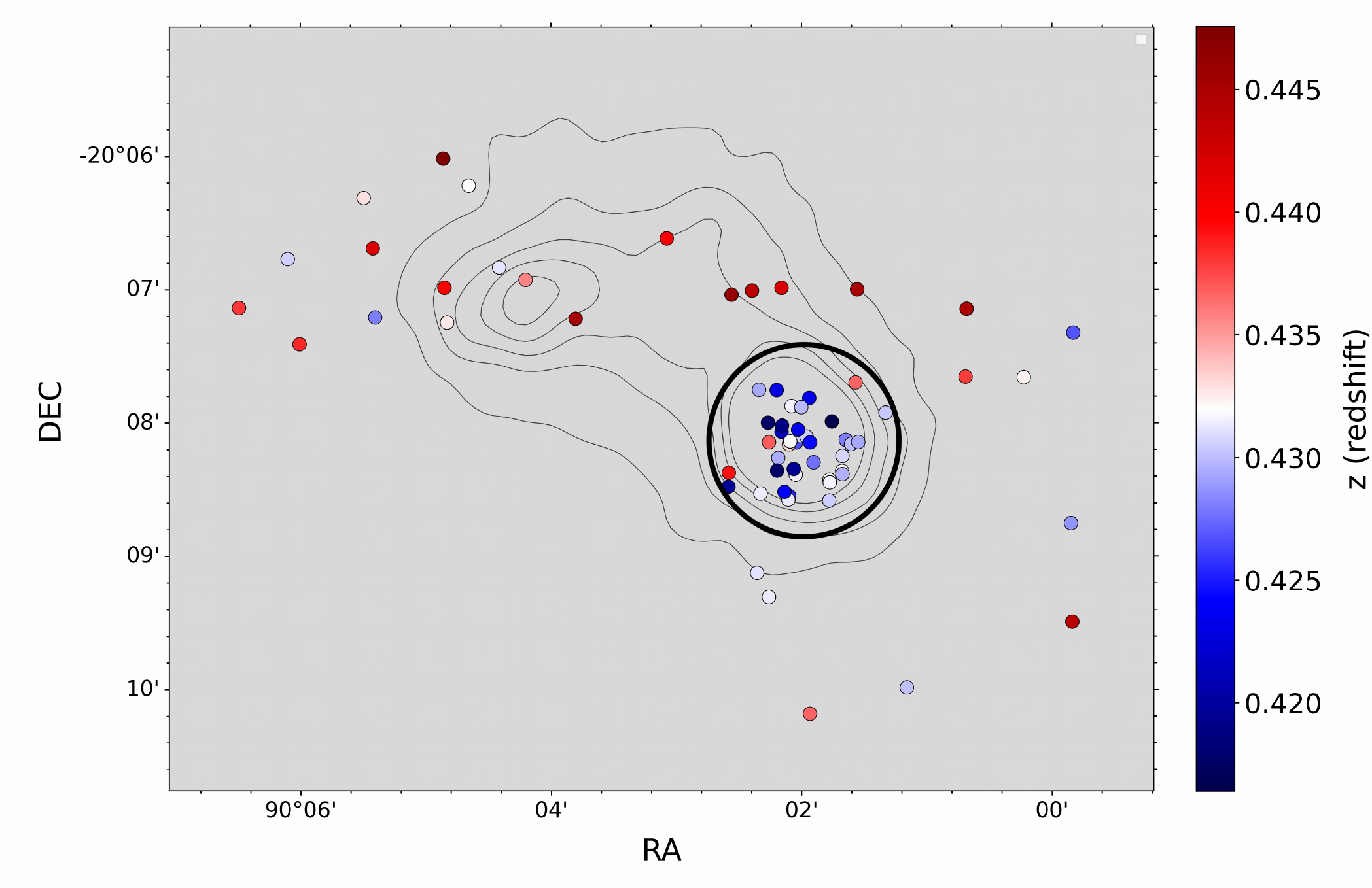}}
\caption{Member galaxy map, with X-ray emission contours. The galaxies reported are the ones with the redshift spectroscopically confirmed and are colored according to the colorbar based on the value of their redshift. The black circle marks the galaxy population associated with the central X-ray peak, corresponding to the region adopted in the KS-test analysis.}
\label{fig:ottico}
\end{figure}

To further support this hypothesis, additional insights into this region can be obtained from the list of spectroscopically confirmed redshifts of 67 cluster members. In Fig.~\ref{fig:ottico}, each galaxy is colored according to the redshift, with the bluer ones having smaller $z$ values ($z\sim 0.420$) and the redder ones having larger ones ($z\sim 0.445$).  
Thanks to the X-ray emission contours, one can notice how the galaxies located close to the cold and X-bright gas core are generally bluer, that is lower $z$, compared to all the other galaxies scattered throughout the rest of the cluster. To verify this statistically, we performed a two-sample Kolmogorov-Smirnov test (KS test) on the two collections of galaxies composed by those contained inside (39 galaxies) and outside (28 galaxies) of the black circle of Fig.~\ref{fig:ottico}.
Considering a confidence level of $95\%$, we can reject the null hypothesis since the p-value obtained is $p\ll0.05$. This indicates that the galaxies within the cooler sub-clump have on average a significant lower redshift, that suggests a relative motion along the line-of-sight toward us with respect to the rest of the cluster.

We have thus compared all the available multiwavelength data for the southern region, from which it emerges that a compact and cooler core, with a peaked X-ray emission, is moving away from the central region, likely producing a cold front ahead of it.

\subsection{Central region}\label{sec:central}
\subsubsection{Analysis of the ICM properties}
The second region we considered, labeled as "Center" in Fig.~\ref{fig:global_view} panel (a), is bright in both X-ray and radio bands. However, a clear offset is observed between the bulk of the radio emission, which is concentrated toward the cluster center, and the X-ray peak, located to the south. In the central part, the X-ray emission appears more diffuse and flatter compared to the southern peak, lacking a pronounced maximum. This same area also hosts the majority of the diffuse radio emission. 

In this same region, the ICM reaches its highest temperature. In Tab.~\ref{tab:temp} one of the measurements we made is reported, indicating a temperature of about $11\hbox{ keV}$. However, several substructures are visible, and in some locations the temperature may be even higher. 

The presence of very hot ICM corresponding to an asymmetric X-ray emission cospatial with a radio halo, are clear symptoms of a recent merger event \citep[e.g.,][]{Cassano2006, Donnert2013}. 
The ICM was likely heated by an energetic event that injected turbulence in the ICM, eventually producing the observed diffuse radio emission. 

We note how the diffuse radio emission is well aligned with the X-ray surface brightness of this region, both in terms of total extension and shape of the sources.
Such a remarkable similarity strongly suggests a common origin of the morphological properties observed in the two bands.
In addition, thanks to the high-sensitivity and resolution of \meerkat{}, we can identify cospatial substructures in the radio and X-ray images. This evidence supports the presence of a local connection among the different plasma components of the ICM.

\subsubsection{Substructure detection}
\begin{figure}
\centering
\resizebox{1\columnwidth}{!}{\includegraphics[]{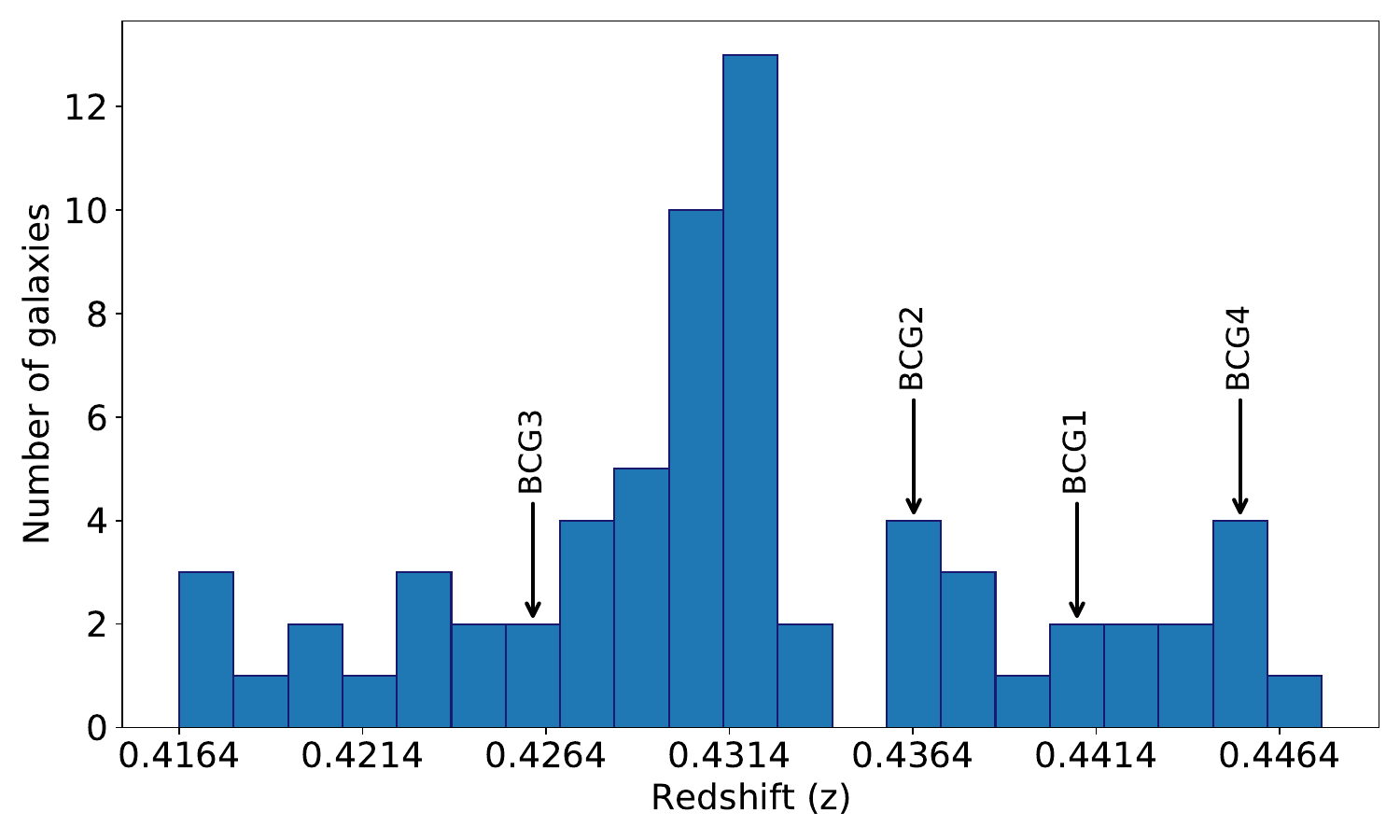}}
\caption{Histogram of all the available spectroscopically confirmed redshifts. The arrows indicate the four BCGs identified by \citet{Furtak2024} in the central and southern region of the cluster.}
\label{fig:isto}
\end{figure}

An additional interesting feature in the central region emerges from the analysis of the redshifts of the cluster member galaxies. Figure~\ref{fig:isto} shows the redshift histogram of the 67 spectroscopically confirmed galaxies  described in the previous section. Among these, \citet{Furtak2024} identify up to three different BCGs in the central part of the system (BCG1, BCG2, BCG4) and one in the southern part (BCG3), each potentially associated with a distinct substructure. In the histogram, the arrows indicate all four BCGs shown and labeled in panel (d) of Fig.~\ref{fig:global_view}. Although the sample available in the central region is more limited than that associated with the southern part of the cluster, the galaxies are nevertheless distributed sufficiently uniformly across the region of interest to allow a reasonably robust analysis to detect the presence of substructures. 

\begin{figure*}
\centering
\begin{minipage}{0.49\textwidth}
    \centering
    \includegraphics[width=\textwidth]{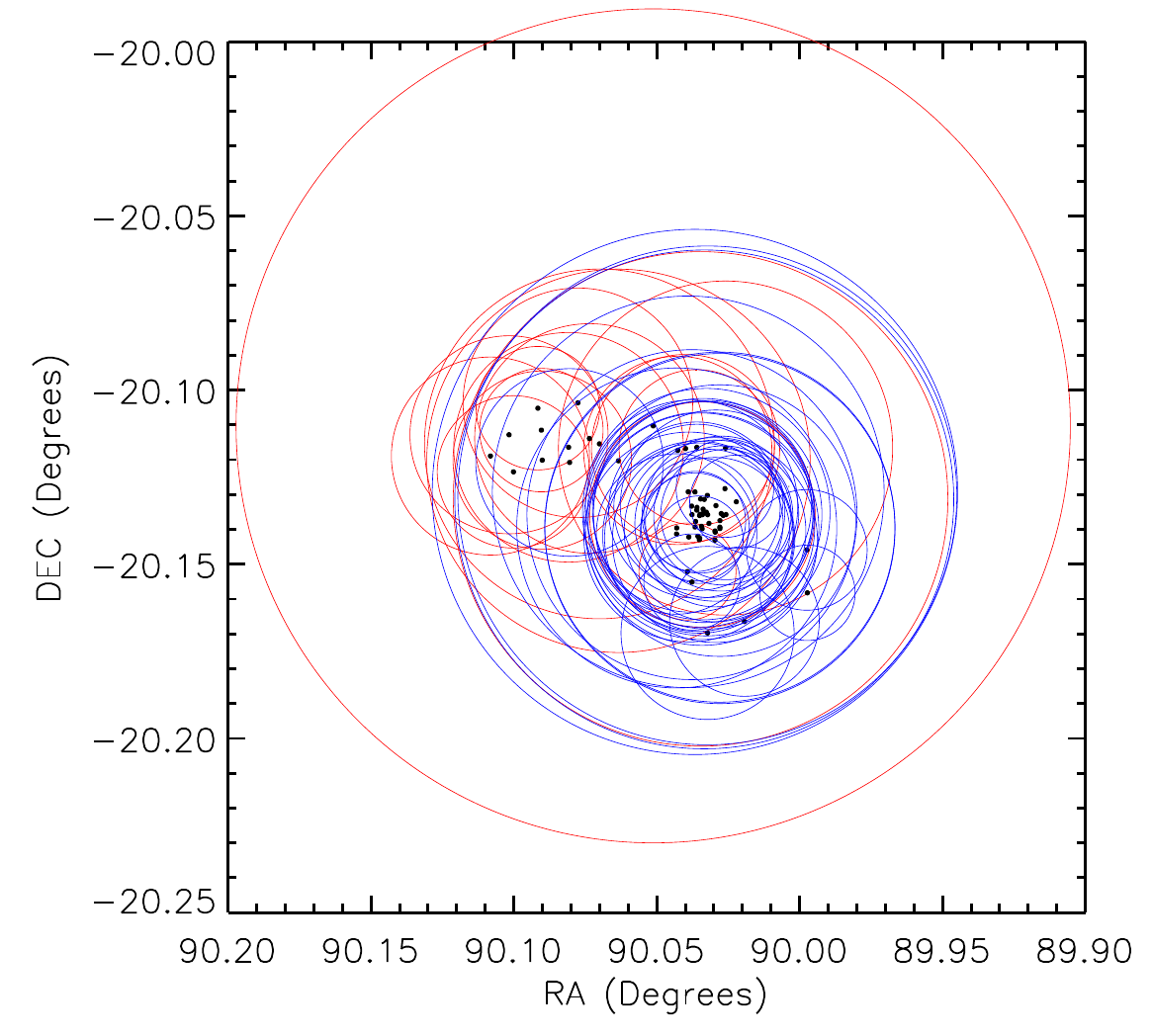}
\end{minipage}
\hfill
\begin{minipage}{0.49\textwidth}
    \centering
    \includegraphics[width=\textwidth]{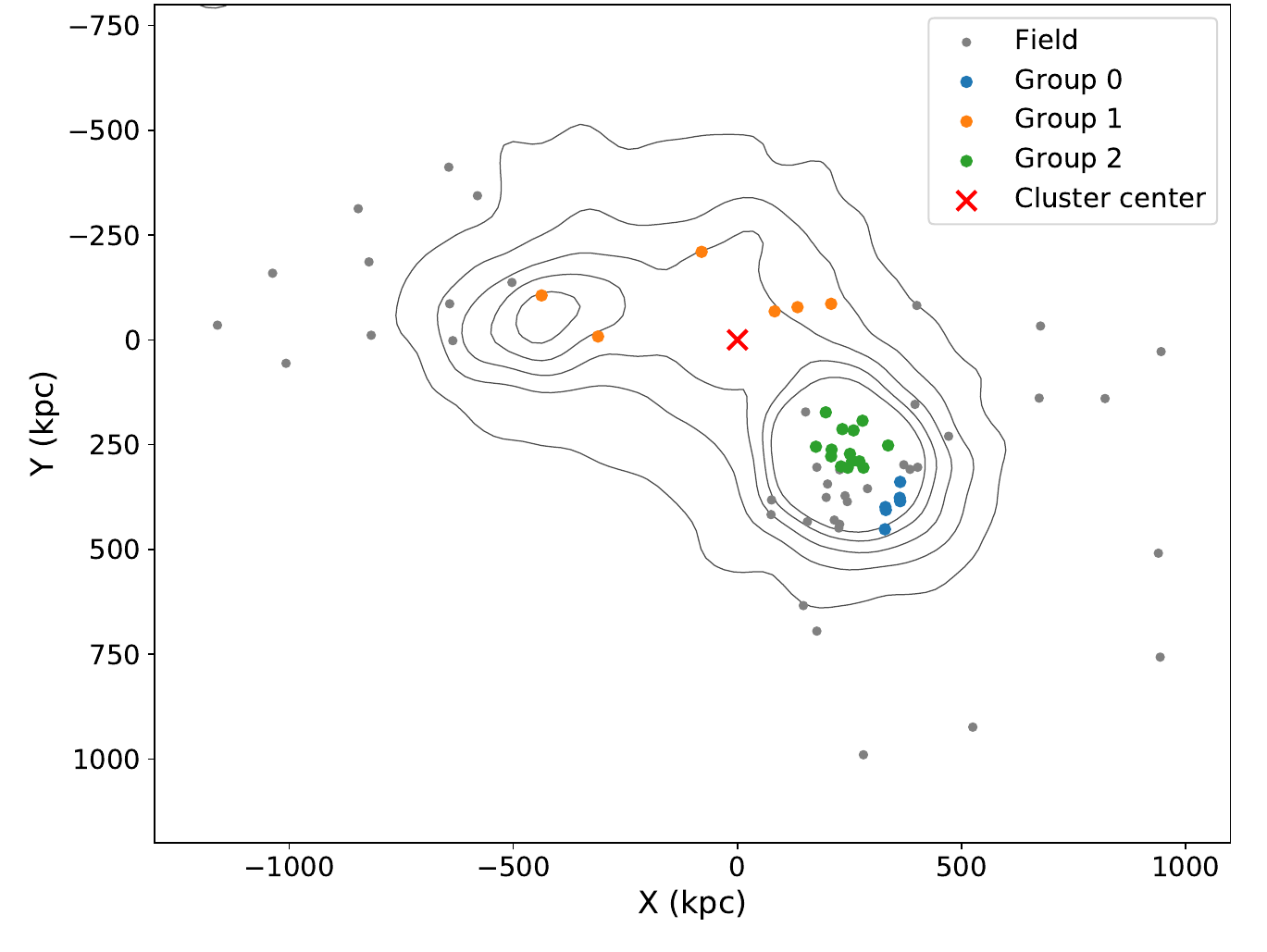}
\end{minipage}
\caption{Results for the three-dimensional tests for substructures. Left: results from the DS substructure test. Each point represents a galaxy, while the size of the surrounding circle is proportional to the local deviation of the velocity distribution with respect to the global cluster kinematics, quantified by $\delta_i$ (see Eq.~\ref{eq:delta_i}). Clustering of large circles indicates significant departures and the likely presence of a substructure. Circle colors are used only to improve the readability of the plot: blue circles correspond to galaxies located in the southern region, while red circles indicate galaxies in the central region of the cluster. Right: results from the DS+ substructure test. Galaxies are colored by DS+ group assignment, while galaxies not associated with any detected substructure are shown in gray. Coordinates are given in projected physical units (kpc) relative to the cluster center.}
\label{fig:bubble}
\end{figure*}

First, we performed one-dimensional tests for the possible deviations from a Gaussian distribution, which can be attributed to dynamical activity. In particular, we converted the redshift of each galaxy into a line-of-sight velocity using the standard special relativistic formula
\begin{equation}
    v_{los}= c\frac{(1+z)^2-1}{(1+z)^2+1}\ ,
    \label{eq:v_los}
\end{equation}
where $c$ is the speed of light. 
In dynamically relaxed clusters, the BCG is expected to have a negligible peculiar velocity with respect to the systemic velocity of the cluster. Significant peculiar velocities may instead indicate a dynamically disturbed system, possibly associated with a recent or ongoing merger event. The peculiar velocities of the BCGs in our sample are: BCG1 = 1736 km/s, BCG2 = 778 km/s, BCG3 = $-$1133 km/s, BCG4 = 2925 km/s. 

Then, we analyzed the entire velocity distribution through the Anderson–Darling (AD) test to detect significant deviations from normality \citep{Hou2009}. We used the AD test as implemented in the task \texttt{ad.test} of the package \texttt{nortest} in version 1.0-4 of the R statistical software environment. \footnote{\hyperlink{http://www.r-project.org/}{http://www.r-project.org/}} 
The Anderson–Darling test yields a p-value of 0.012, which is generally regarded as significant for the detection of departures from a single gaussian \citep[e.g.,][]{Einasto2012}, even though the significance of the test is below $3\sigma$.

As an additional one-dimensional consistency check for the possible presence of substructures, we applied the Gaussian mixture modeling approach implemented in the \texttt{mclust} package \citep{Scrucca2016}.\footnote{\hyperlink{https://mclust-org.github.io/mclust/}{https://mclust-org.github.io/mclust/}} 
Testing models with up to 9 Gaussian components, the Bayesian Information Criterion (BIC) slightly favors a three-component solution over a single Gaussian, while two-component models are not preferred. However, the BIC difference between the one- and three-component models is $\Delta\mathrm{BIC} \sim 1$, well below the threshold of 10 commonly adopted to consider multimodality as significant \citep{Kass1995}. Therefore, this test alone provides no robust evidence for distinct kinematic substructures in the line-of-sight velocity distribution.

In any case, one-dimensional tests, based solely on the line-of-sight velocity distribution, are less sensitive to physically meaningful substructures than three-dimensional tests, which combine velocity and projected position information. In particular, we note that the spectroscopic sample is not spatially uniform across the cluster, with a significant fraction of redshift measurements concentrated in the X-ray peak region. The one-dimensional velocity distribution is therefore likely biased by this sampling inhomogeneity, as the prominent peak in the redshift histogram may reflect the spatial coverage of our observations rather than a genuine kinematic substructure. Indeed, incorporating the spatial distribution of galaxies is essential to identify and physically interpret distinct kinematic subgroups within the cluster \citep{Pinkney1996, Gastaldello2013}. 

As a first approach we adopt the Dressler–Shectman (DS) $\Delta$ statistic \citep{DS1988} which tests for differences in the local mean and dispersion compared to the global mean and dispersion of the cluster. The local velocity anisotropy is calculated for each galaxy as 
\begin{equation}
    \delta^2=\bigg(\frac{N_{nn}+1}{\sigma^2}\bigg)\ [(\bar{v}_{loc}-\bar{v})^2+(\sigma_{loc}-\sigma)^2] ,
    \label{eq:delta_i}
\end{equation}
where $N_{nn}$ is the number of nearest neighbors used to compute the local mean recession velocity ($\bar{v}_{loc}$) and local velocity dispersion ($\sigma_{loc}$), while $\bar{v}$ and $\sigma$ are the global mean velocity and velocity dispersion of the cluster. Following \citet{Pinkney1996}, we adopt $N_{nn} = \sqrt{N}$, where $N$ is the total number of spectroscopic members. The calculation of the $\Delta$ statistic involves the summation of these local velocity anisotropies. The significance of the $\Delta$ statistic is estimated via 10000 Monte Carlo simulations in which galaxy velocities are randomly shuffled while positions are kept fixed. 
We obtain a total $\Delta = 124.8$, significantly larger than the mean value expected from the simulations ($\langle \Delta \rangle = 83.2 \pm 11.2$). None of the randomized realizations produced a value larger than the observed one, corresponding to a probability of chance occurrence $P \simeq 9.8 \times 10^{-5}$. This indicates a highly significant detection of substructure within the cluster. 
The results are visualized in the bubble plot in Fig.~\ref{fig:bubble}, where each galaxy is represented by a circle whose size is proportional to $e^{\delta_i}$. Clustering of large circles indicates significant local departures from the global kinematics, and hence the likely presence of substructure. One can notice how, besides several large circles concentrated around the southern X-ray peak, a number of medium-to-large circles are found on either side of the central region, suggesting possible kinematic departures from the global cluster dynamics in these areas. These results further suggests that the cluster may be undergoing an advanced stage of a multiple merger.

We further tested for the presence of subgroups in addition to the southern clump using the DS+ method \citep{Benavides2023},\footnote{\hyperlink{https://github.com/josegit88/MilaDS}{https://github.com/josegit88/MilaDS}} an extension of the classical Dressler-Shectman test that not only assesses the statistical significance of substructure but also identifies individual subgroups and assigns member galaxies to each of them. Unlike the original DS test, which provides a single global significance estimate, DS+ iteratively searches for galaxy groups in the combined position-velocity space. 
The DS+ algorithm requires as input a reference cluster redshift and a cluster center to define the global kinematic properties of the system. For both quantities, we adopted the values reported in literature, namely $z_{cl} = 0.43$ and $\mathrm{RA} = 06^{\mathrm h}00^{\mathrm m}11.3^{\mathrm s}$, $\mathrm{Dec} = -20^\circ07^\prime14.5^{\prime\prime}$ (J2000) for the cluster center. 
Additionally, it considers the peculiar velocities of each galaxy, calculated using Eq.~\ref{eq:v_los} with the peculiar redshift replacing z, computed as
\begin{equation}
    z_{spec}= \frac{z-z_{cl}}{1+z_{cl}}.
\end{equation}
As visible in Fig.~\ref{fig:bubble}, our DS+ analysis identifies three significant subgroups: the first two are spatially concentrated around the southern X-ray peak, confirming the presence of a distinct kinematic clump in that region; the third group is centered near BCG4 in the central region of the cluster, suggesting a further kinematically distinct component.
Notably, galaxies in the vicinity of BCG1 and BCG2 are not assigned to any subgroup, indicating that their kinematics are consistent with the global velocity distribution of the main cluster.

To summarize all the results of the substructure detection tests, we find compelling evidence for a dynamically complex system, characterized by multiple kinematically distinct components in the central and southern regions. These results strongly support the scenario of an ongoing multi-component merger, in which several substructures are still actively interacting within the cluster core.

\subsection{North region}\label{sec:north}
The third and final region is labeled as "North" in Fig.~\ref{fig:global_view} panel (a). 
In analyzing this region, we also include the northern structures that \citet{Furtak2024} incorporate in their improved strong lensing model of its mass distribution. Here one can notice two X-ray emission clumps probably lying at the same redshift as the rest of the cluster. Unfortunately, there is not enough  X-ray lines statistics available to determine their redshift. 
Therefore, all spectral quantities derived for this region, including temperature and metallicity, were obtained by assuming the same redshift as the main cluster. If the actual redshift differs from the adopted value, the inferred thermodynamic properties may be affected by systematic biases. 
However, \citet{Furtak2024} considers them part of the cluster because of the observed overdensity of probable cluster galaxies in that area. 

The gas temperature is found to be quite low, as reported in Tab.~\ref{tab:temp}: indeed the lowest temperature reached in this region is about $2\hbox{ keV}$. 
These cold gas structures probably are not new material accreted from a filament of the cosmic web because their metallicity ($0.49_{-0.17}^{+0.18}\ Z_{\odot}$) is consistent with the value in the central region, always around $0.3Z_{\odot}$. This indicates that it is already processed material, probably belonging to two small groups now infalling onto the central cluster. 
The lack of detection of radio emission in these region further support the fact that these structures are still in the early phases of accretion on the main cluster. In fact, a diffuse, radio inter-cluster emission would be expected both during the merging phase or in a later stage \citep{Nishiwaki2026} as observed in past cases \citep{Pignataro2024,Hu2025}.

\begin{figure}
\centering
\resizebox{0.8\columnwidth}{!}{\includegraphics[]{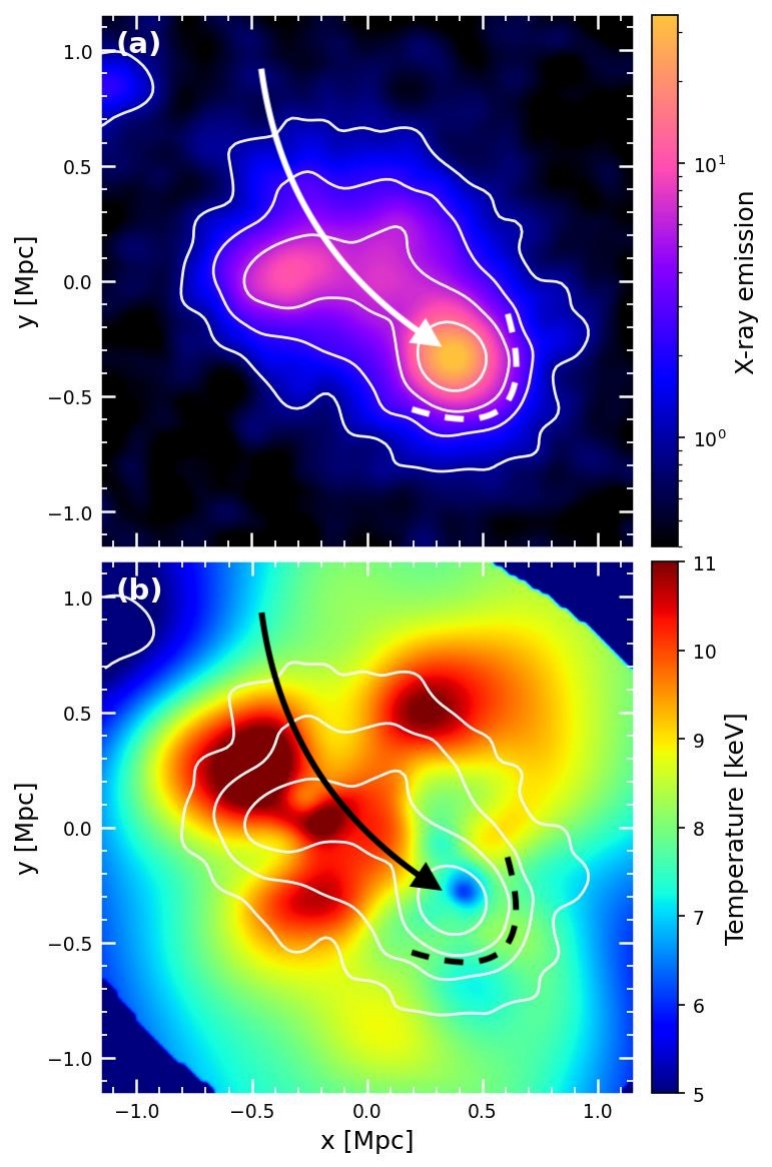}}
\caption{Enlargement of the X-ray emission (a) and temperature (b) maps on the central region, with X-ray emission contours. Dotted in black is the position of the SB discontinuity of the X-ray emission, while the arrow indicates the motion we hypothesize for the cool core through the main cluster.}
\label{fig:confronto_Lysk}
\end{figure}

\section{Results and discussion}\label{sec:sec4}
The emerging picture from the previous section is a complex merger dynamics scenario. Our analysis reveals the presence of a compact gas clump, which exhibits a sharply peaked and nearly symmetric emission profile, that has retained a lower temperature than the surrounding medium in which it is moving. In particular, it appears to have already crossed the central region, heating the gas to very high temperatures, driving turbulence in the intracluster medium, and contributing to produce the diffuse radio emission. This motion has a significant component along the line of sight, as indicated by the redshift distribution of the member galaxies, and likely also a component toward the south, giving rise to the observed X-ray surface brightness discontinuity in that direction. 
This could mean that a compact cool core cluster crossed a second, more massive cluster while maintaining its core of cold, X-ray bright gas. In contrast, the cluster that has been crossed has completely lost its ordered structure and exhibits a highly disturbed morphology. 

To further investigate this scenario, we compared our findings with previous numerical simulations of cluster mergers. In particular, \citet{Lyskova2019} studied the interaction between the Coma cluster and the NGC 4839 group, reproducing the gas surface brightness and temperature distributions throughout the merger. Considering the panels (2a) and (2b) of Fig.~8 of the aforementioned paper, we compared this stage of the simulated merger with our case (Fig.~\ref{fig:confronto_Lysk}). While the Coma cluster provides a useful reference, MACS0600 shows some differences: the cool core has not yet reversed its motion, and the infalling system likely has a higher mass ratio relative to the main cluster, producing a stronger impact on the central gas. Despite these differences, the overall configuration -- disturbed central X-ray morphology, elevated core temperature, and a compact cool core associated with a cold front -- supports a post-merger interpretation for MACS0600, consistent with the qualitative features seen in these simulations.

\begin{figure}
\centering
\resizebox{1\columnwidth}{!}{\includegraphics[]{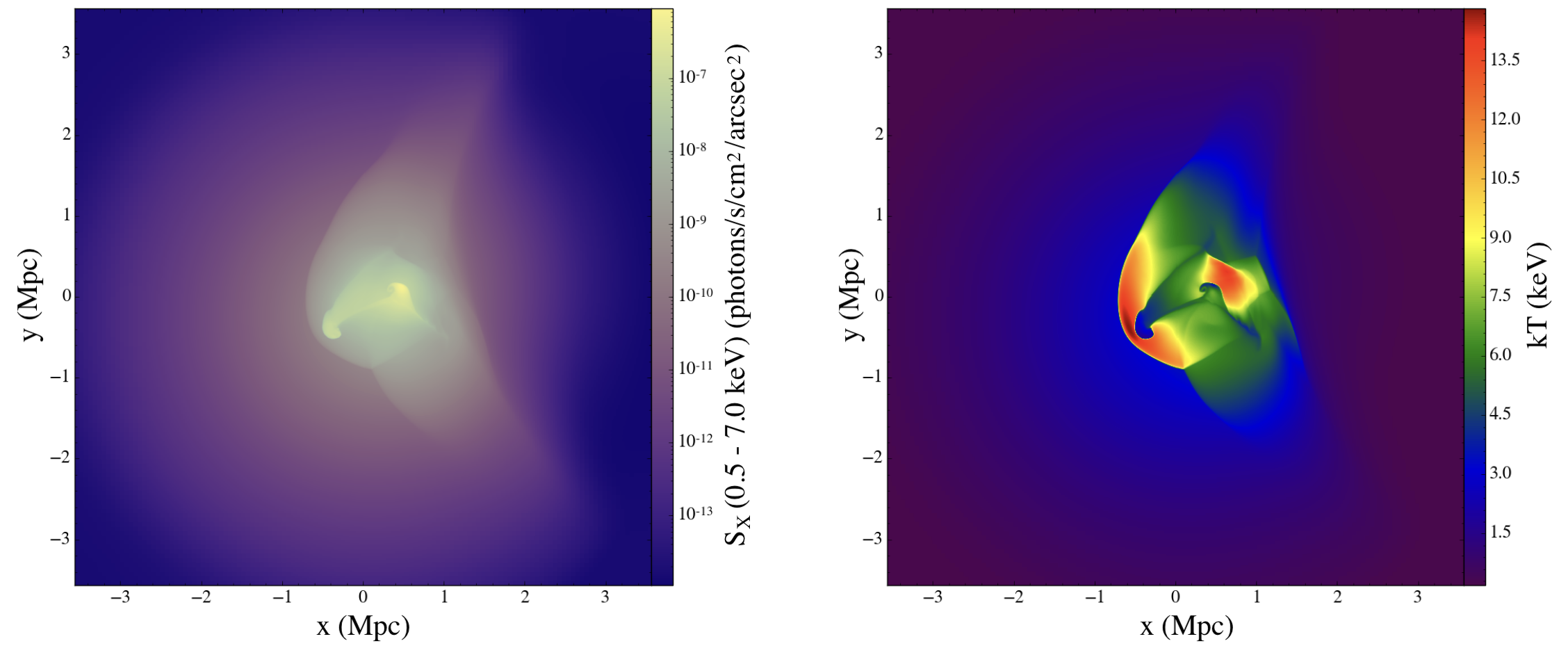}}
\caption{Simulation result for a binary merger with mass ratio between the two clusters $R=1:3$ and initial impact parameter $b=500\hbox{ kpc}$. X-ray emissivity (on the left) and gas temperature (on the right). Achieved thanks to the simulation tool of \citet{ZuHone2011}.}
\label{fig:simulazione}
\end{figure}

To verify that a similar result can also be obtained in cases other than the specific one of the Coma cluster, we used the data from the Galaxy Cluster Merger Catalog,\footnote{\hyperlink{http://gcmc.hub.yt}{http://gcmc.hub.yt}} a collection of mock observations from simulated galaxy cluster mergers \citep{ZuHone2018}. In particular, we observed the simulation outcomes gathered in the section \textit{A Parameter Space Exploration of Galaxy Cluster Mergers}, which refers to the study of \citet{ZuHone2011}. One can then visualize the simulation outcomes for different mass ratio ($R$) between the two clusters and different initial impact parameter ($b$) in a binary merger. For instance, for $R=1:3$ and $b=500\hbox{ kpc}$ the simulated scenario produces features quite similar to the ones of MACS0600. Indeed, in Fig.~\ref{fig:simulazione}, one can see how the central ICM is heated and its morphology disturbed. In the meanwhile, the bullet remains bright in the X-rays, cold, and it produces a cold front by moving inside a hotter gas. This suggests that, although the parameters adopted in the simulation do not exactly match those best representing MACS0600, the outcome of a binary merger -- in which a compact cool core goes through a more massive cluster, perturbing it while remaining relatively intact -- closely resembles what is observed in the system analyzed in this work. Thus, the analysis of the data presented in this paper did allow us to provide a possible merging scenario for MACS0600, also supported by the numerical simulations.

While the present interpretation provides a robust framework for understanding the system, future studies could clarify whether it represents only a first-order description and whether the cluster is undergoing a more complex, multi-component merger. 
In particular, a denser spectroscopic coverage of cluster member galaxies across the entire system would allow a more comprehensive optical substructure analysis, improving the identification and characterization of kinematically distinct components. 
In addition, a more detailed study of the central region, including radio spectral and polarization studies to better characterize the non-thermal emission, could improve our understanding of the gas substructures and their dynamics. 

Additionally, if deeper X-ray observations of the southern region become available in the future, they could allow a better characterization of the cold front, for instance through a measurement of the temperature jump.

\section{Summary and Conclusions}\label{sec:conclusions}
In this work, we investigated the merger dynamics of the massive and complex cluster MACS0600 using a multi-wavelength approach. We analyzed the cluster morphology and the thermodynamic properties of the ICM through \xmm{} and \chandra{} X-ray observations, and explored the non-thermal component via diffuse radio emission observed with \meerkat{}. Our main results can be summarized as follows:
\begin{itemize}
\item the cluster exhibits a complex morphology, with multiple substructures and emission clumps in addition to the main X-ray peak. A clear offset is observed between the bulk of the radio emission and the X-ray peak;
\item at the X-ray peak, a gas core is detected with a temperature lower than that of the surrounding medium ($\sim6\pm1\ \mathrm{keV}$ compared to $\sim10\pm3\ \mathrm{keV}$). A sharp surface brightness discontinuity is observed around this region, strongly suggesting the presence of a cold front, although the current data do not allow us to constrain the corresponding temperature jump;
\item the central region reaches temperatures of $\sim10\ \mathrm{keV}$, higher than in the outer regions, likely due to shock heating and turbulence induced by the merger. Most of the diffuse radio emission is located in this area, supporting a scenario in which merger-driven turbulence both heats the gas and generates synchrotron emission. A faint radio tail appears to connect the central halo with the X-ray peak, tracing its motion;
\item the relative motion between the central region and the compact cool core is supported by optical data from \citet{Furtak2024}. We find that galaxies associated with the cool clump have, on average, lower redshifts than the rest of the cluster members, indicating motion along the line of sight toward the observer. This is consistent with a scenario in which the cool core is moving away from the cluster center after its passage through it;
\item the presence of kinematically distinct substructures is confirmed by statistical tests such as the DS and DS+ methods;
\item comparison with numerical simulations of cluster mergers, in particular \citet{Lyskova2019} and \citet{ZuHone2018}, supports a post-merger scenario and provides a consistent interpretation of the observed features.
\end{itemize}
We conclude that MACS0600 is undergoing a merger in which a compact cool core has crossed the main, more massive cluster without being completely disrupted, while significantly perturbing the surrounding ICM. This would explain the motion away from the central zone of the cold clump, which, still retains a peaked and almost symmetrical X-ray morphology. Although the core remains cooler than the surrounding gas, it is hotter than typical cool cores, likely due to heating during the interaction. The merger also naturally accounts for the turbulent state of the ICM and the associated diffuse radio emission.

This system represents a rare example of such a merger configuration. While many merging clusters have been identified in large surveys, only a limited number have been studied in detail through multiwavelength analyses and dynamical reconstruction. MACS0600 therefore provides a valuable laboratory for investigating the physics of cluster mergers and the evolution of the ICM. This single case study is representative of a region of the parameter space of galaxy cluster mergers with a combination of mass ratio and impact parameter in which a subcluster is able to preserve a remarkably compact and X-ray-bright core, highlighting the resilience of dense gas structures during the interaction. It also shows that the location of the X-ray peak does not necessarily coincide with that of the dominant mass concentration, emphasizing the importance of distinguishing between the distribution of the emitting gas and the underlying mass. The identification of discontinuities, such as cold fronts (Fig.~\ref{fig:jump}), further provides a powerful diagnostic of the dynamical history of the system, as these structures can retain signatures of the merger (e.g., the direction of motions and the subcluster velocities) well after the initial interaction. 
In the central region, the spatial morphology and location of the diffuse radio emission further supports a scenario in which merger-driven turbulence re-accelerates relativistic electrons, possibly through an inefficient process requiring time to develop. Taken together, these observations provide a consistent picture of the interplay between merger dynamics, ICM heating, turbulence, and non-thermal emission. 
Most importantly, our results demonstrate that the dynamical state of a merging cluster cannot be fully understood from a single wavelength alone: only through the combined analysis of the X-ray, radio and optical emission does the complex merger geometry and its associated physical processes emerge clearly. In this sense, multiwavelength observations are essential for disentangling the mechanisms shaping the evolution of merging galaxy clusters.\\
\\
\begin{acknowledgements}
MB acknowledge support from the ERC CoG $\vec{B}$ELOVED, GA n. 101169773. 
MB, FG, MR acknoledge the financial contribution from the INAF GO grant 1.05.24.02.10 Extended Radio Emission in Galaxy Clusters at deep focus with MeerKAT. 
The MeerKAT telescope is operated by the South African Radio Astronomy Observatory, which is a facility of the National Research Foundation, an agency of the Department of Science and Innovation.
\end{acknowledgements}

\bibliography{bib}

@ARTICLE{arnaud2010,
       author = {{Arnaud}, M. and {Pratt}, G.~W. and {Piffaretti}, R. and {B{\"o}hringer}, H. and {Croston}, J.~H. and {Pointecouteau}, E.},
        title = "{The universal galaxy cluster pressure profile from a representative sample of nearby systems (REXCESS) and the Y$_{SZ}$ - M$_{500}$ relation}",
      journal = {\aap},
         year = 2010,
        month = jul,
       volume = {517},
          eid = {A92},
        pages = {A92},
          doi = {10.1051/0004-6361/200913416},
archivePrefix = {arXiv},
       eprint = {0910.1234},
 primaryClass = {astro-ph.CO},
       adsurl = {https://ui.adsabs.harvard.edu/abs/2010A&A...517A..92A}
}

@ARTICLE{kravtsov2012,
       author = {{Kravtsov}, Andrey V. and {Borgani}, Stefano},
        title = "{Formation of Galaxy Clusters}",
      journal = {\araa},
         year = 2012,
        month = sep,
       volume = {50},
        pages = {353-409},
          doi = {10.1146/annurev-astro-081811-125502},
archivePrefix = {arXiv},
       eprint = {1205.5556},
 primaryClass = {astro-ph.CO},
       adsurl = {https://ui.adsabs.harvard.edu/abs/2012ARA&A..50..353K}}

@ARTICLE{2006Croston,
       author = {{Croston}, J.~H. and {Arnaud}, M. and {Pointecouteau}, E. and {Pratt}, G.~W.},
        title = "{An improved deprojection and PSF-deconvolution technique for galaxy-cluster X-ray surface-brightness profiles}",
      journal = {\aap},
         year = 2006,
        month = dec,
       volume = {459},
       number = {3},
        pages = {1007-1019},
          doi = {10.1051/0004-6361:20065795},
archivePrefix = {arXiv},
       eprint = {astro-ph/0608700},
 primaryClass = {astro-ph},
       adsurl = {https://ui.adsabs.harvard.edu/abs/2006A&A...459.1007C}
}

@ARTICLE{Bartalucci2017,
       author = {{Bartalucci}, I. and {Arnaud}, M. and {Pratt}, G.~W. and {Vikhlinin}, A. and {Pointecouteau}, E. and {Forman}, W.~R. and {Jones}, C. and {Mazzotta}, P. and {Andrade-Santos}, F.},
        title = "{Recovering galaxy cluster gas density profiles with XMM-Newton and Chandra}",
      journal = {\aap},
         year = 2017,
        month = dec,
       volume = {608},
          eid = {A88},
        pages = {A88},
          doi = {10.1051/0004-6361/201731689},
archivePrefix = {arXiv},
       eprint = {1709.06570},
 primaryClass = {astro-ph.CO},
       adsurl = {https://ui.adsabs.harvard.edu/abs/2017A&A...608A..88B}
}

@INPROCEEDINGS{Rahaman2020,
       author = {{Rahaman}, M. and {Datta}, A. and {Raja}, R.},
        title = "{Investigating the origin of diffuse radio emission in galaxy clusters {\textemdash} using radio and X-ray observations}",
    booktitle = {American Astronomical Society Meeting Abstracts \#236},
         year = 2020,
       series = {American Astronomical Society Meeting Abstracts},
       volume = {236},
        month = jun,
          eid = {127.02},
        pages = {127.02},
       adsurl = {https://ui.adsabs.harvard.edu/abs/2020AAS...23612702R}}

@article{vanWeeren2019,
       author = {{van Weeren}, R.J. and {de Gasperin}, F. and {Akamatsu}, H. and {Br{\"u}ggen}, M. and {Feretti}, L. and {Kang}, H. and {Stroe}, A. and {Zandanel}, F.},
        title = "{Diffuse Radio Emission from Galaxy Clusters}",
      journal = {\ssr},
         year = 2019,
        month = feb,
       volume = {215},
       number = {1},
          eid = {16},
        pages = {16},
          doi = {10.1007/s11214-019-0584-z},
archivePrefix = {arXiv},
       eprint = {1901.04496},
 primaryClass = {astro-ph.HE},
       adsurl = {https://ui.adsabs.harvard.edu/abs/2019SSRv..215...16V}}

@ARTICLE{adam2017_a,
       author = {{Adam}, R. and {Bartalucci}, I. and {Pratt}, G.~W. and {Ade}, P. and {Andr{\'e}}, P. and {Arnaud}, M. and {Beelen}, A. and {Beno{\^\i}t}, A. and {Bideaud}, A. and {Billot}, N. and {Bourdin}, H. and {Bourrion}, O. and {Calvo}, M. and {Catalano}, A. and {Coiffard}, G. and {Comis}, B. and {D'Addabbo}, A. and {De Petris}, M. and {D{\'e}mocl{\`e}s}, J. and {D{\'e}sert}, F.-X. and {Doyle}, S. and {Egami}, E. and {Ferrari}, C. and {Goupy}, J. and {Kramer}, C. and {Lagache}, G. and {Leclercq}, S. and {Mac{\'\i}as-P{\'e}rez}, J.-F. and {Maurogordato}, S. and {Mauskopf}, P. and {Mayet}, F. and {Monfardini}, A. and {Mroczkowski}, T. and {Pajot}, F. and {Pascale}, E. and {Perotto}, L. and {Pisano}, G. and {Pointecouteau}, E. and {Ponthieu}, N. and {Rev{\'e}ret}, V. and {Ritacco}, A. and {Rodriguez}, L. and {Romero}, C. and {Ruppin}, F. and {Schuster}, K. and {Sievers}, A. and {Triqueneaux}, S. and {Tucker}, C. and {Zemcov}, M. and {Zylka}, R.},
        title = "{Mapping the kinetic Sunyaev-Zel'dovich effect toward MACS J0717.5+3745 with NIKA}",
      journal = {\aap},
         year = 2017,
        month = feb,
       volume = {598},
          eid = {A115},
        pages = {A115},
          doi = {10.1051/0004-6361/201629182},
archivePrefix = {arXiv},
       eprint = {1606.07721},
 primaryClass = {astro-ph.CO},
       adsurl = {https://ui.adsabs.harvard.edu/abs/2017A&A...598A.115A} }

@article{Ebeling2001,
doi = {10.1086/320958},
url = {https://doi.org/10.1086/320958},
year = {2001},
month = {jun},
publisher = {},
volume = {553},
number = {2},
pages = {668},
author = {Ebeling, H. and Edge, A. C. and Henry, J. P.},
title = {MACS: A Quest for the Most Massive Galaxy Clusters in the Universe},
journal = {The Astrophysical Journal}}

@article{CHEXMATE2021,
       author = {{CHEX-MATE Collaboration} and {Arnaud}, M. and {Ettori}, S. and {Pratt}, G.~W. and {Rossetti}, M. and {Eckert}, D. and {Gastaldello}, F. and {Gavazzi}, R. and {Kay}, S.~T. and {Lovisari}, L. and {Maughan}, B.~J. and {Pointecouteau}, E. and {Sereno}, M. and {Bartalucci}, I. and {Bonafede}, A. and {Bourdin}, H. and {Cassano}, R. and {Duffy}, R.~T. and {Iqbal}, A. and {Maurogordato}, S. and {Rasia}, E. and {Sayers}, J. and {Andrade-Santos}, F. and {Aussel}, H. and {Barnes}, D.~J. and {Barrena}, R. and {Borgani}, S. and {Burkutean}, S. and {Clerc}, N. and {Corasaniti}, P. -S. and {Cuillandre}, J. -C. and {De Grandi}, S. and {De Petris}, M. and {Dolag}, K. and {Donahue}, M. and {Ferragamo}, A. and {Gaspari}, M. and {Ghizzardi}, S. and {Gitti}, M. and {Haines}, C.~P. and {Jauzac}, M. and {Johnston-Hollitt}, M. and {Jones}, C. and {K{\'e}ruzor{\'e}}, F. and {Le Brun}, A.~M.~C. and {Mayet}, F. and {Mazzotta}, P. and {Melin}, J. -B. and {Molendi}, S. and {Nonino}, M. and {Okabe}, N. and {Paltani}, S. and {Perotto}, L. and {Pires}, S. and {Radovich}, M. and {Rubino-Martin}, J. -A. and {Salvati}, L. and {Saro}, A. and {Sartoris}, B. and {Schellenberger}, G. and {Streblyanska}, A. and {Tarr{\'\i}o}, P. and {Tozzi}, P. and {Umetsu}, K. and {van der Burg}, R.~F.~J. and {Vazza}, F. and {Venturi}, T. and {Yepes}, G. and {Zarattini}, S.},
        title = "{The Cluster HEritage project with XMM-Newton: Mass Assembly and Thermodynamics at the Endpoint of structure formation. I. Programme overview}",
      journal = {\aap},
         year = 2021,
        month = jun,
       volume = {650},
          eid = {A104},
        pages = {A104},
          doi = {10.1051/0004-6361/202039632},
archivePrefix = {arXiv},
       eprint = {2010.11972},
 primaryClass = {astro-ph.CO},
       adsurl = {https://ui.adsabs.harvard.edu/abs/2021A&A...650A.104C}}

@ARTICLE{Campitiello2022,
       author = {{Campitiello}, M.~G. and {Ettori}, S. and {Lovisari}, L. and {Bartalucci}, I. and {Eckert}, D. and {Rasia}, E. and {Rossetti}, M. and {Gastaldello}, F. and {Pratt}, G.~W. and {Maughan}, B. and {Pointecouteau}, E. and {Sereno}, M. and {Biffi}, V. and {Borgani}, S. and {De Luca}, F. and {De Petris}, M. and {Gaspari}, M. and {Ghizzardi}, S. and {Mazzotta}, P. and {Molendi}, S.},
        title = "{CHEX-MATE: Morphological analysis of the sample}",
      journal = {\aap},
         year = 2022,
        month = sep,
       volume = {665},
          eid = {A117},
        pages = {A117},
          doi = {10.1051/0004-6361/202243470},
archivePrefix = {arXiv},
       eprint = {2205.11326},
 primaryClass = {astro-ph.CO},
       adsurl = {https://ui.adsabs.harvard.edu/abs/2022A&A...665A.117C}}

@ARTICLE{Furtak2024,
       author = {{Furtak}, Lukas J. and {Zitrin}, Adi and {Richard}, Johan and {Eckert}, Dominique and {Sayers}, Jack and {Ebeling}, Harald and {Fujimoto}, Seiji and {Laporte}, Nicolas and {Lagattuta}, David and {Limousin}, Marceau and {Mahler}, Guillaume and {Meena}, Ashish K. and {Andrade-Santos}, Felipe and {Frye}, Brenda L. and {Jauzac}, Mathilde and {Koekemoer}, Anton M. and {Kohno}, Kotaro and {Espada}, Daniel and {Lu}, Harry and {Massey}, Richard and {Niemiec}, Anna},
        title = "{A complex node of the cosmic web associated with the massive galaxy cluster MACS J0600.1-2008}",
      journal = {Monthly Notices of the Royal Astronomical Society},
    volume = {533},
    number = {2},
    pages = {2242-2261},
    year = {2024},
    month = {08},
    issn = {0035-8711},
    doi = {10.1093/mnras/stae1943},
    url = {https://doi.org/10.1093/mnras/stae1943},
    eprint = {https://academic.oup.com/mnras/article-pdf/533/2/2242/58935416/stae1943.pdf}}

@article{Repp2018,
       author = {{Repp}, A. and {Ebeling}, H.},
        title = "{Science from a glimpse: Hubble SNAPshot observations of massive galaxy clusters}",
      journal = {\mnras},
         year = 2018,
        month = sep,
       volume = {479},
       number = {1},
        pages = {844-864},
          doi = {10.1093/mnras/sty1489},
archivePrefix = {arXiv},
       eprint = {1706.01263},
 primaryClass = {astro-ph.GA},
       adsurl = {https://ui.adsabs.harvard.edu/abs/2018MNRAS.479..844R}}

@article{PlanckCollaboration2016,
	author = {{Planck Collaboration} and {Ade, P. A. R.} and {Aghanim, N.} and {Arnaud, M.} and {Ashdown, M.} and {Aumont, J.} and {Baccigalupi, C.} and {Banday, A. J.} and {Barreiro, R. B.} and {Barrena, R.} and {Bartlett, J. G.} and {Bartolo, N.} and {Battaner, E.} and {Battye, R.} and {Benabed, K.} and {Beno\^{\i}t, A.} and {Benoit-L\'evy, A.} and {Bernard, J.-P.} and {Bersanelli, M.} and {Bielewicz, P.} and {Bikmaev, I.} and {B\"ohringer, H.} and {Bonaldi, A.} and {Bonavera, L.} and {Bond, J. R.} and {Borrill, J.} and {Bouchet, F. R.} and {Bucher, M.} and {Burenin, R.} and {Burigana, C.} and {Butler, R. C.} and {Calabrese, E.} and {Cardoso, J.-F.} and {Carvalho, P.} and {Catalano, A.} and {Challinor, A.} and {Chamballu, A.} and {Chary, R.-R.} and {Chiang, H. C.} and {Chon, G.} and {Christensen, P. R.} and {Clements, D. L.} and {Colombi, S.} and {Colombo, L. P. L.} and {Combet, C.} and {Comis, B.} and {Couchot, F.} and {Coulais, A.} and {Crill, B. P.} and {Curto, A.} and {Cuttaia, F.} and {Dahle, H.} and {Danese, L.} and {Davies, R. D.} and {Davis, R. J.} and {de Bernardis, P.} and {de Rosa, A.} and {de Zotti, G.} and {Delabrouille, J.} and {D\'esert, F.-X.} and {Dickinson, C.} and {Diego, J. M.} and {Dolag, K.} and {Dole, H.} and {Donzelli, S.} and {Dor\'e, O.} and {Douspis, M.} and {Ducout, A.} and {Dupac, X.} and {Efstathiou, G.} and {Eisenhardt, P. R. M.} and {Elsner, F.} and {En\ss{}lin, T. A.} and {Eriksen, H. K.} and {Falgarone, E.} and {Fergusson, J.} and {Feroz, F.} and {Ferragamo, A.} and {Finelli, F.} and {Forni, O.} and {Frailis, M.} and {Fraisse, A. A.} and {Franceschi, E.} and {Frejsel, A.} and {Galeotta, S.} and {Galli, S.} and {Ganga, K.} and {G\'enova-Santos, R. T.} and {Giard, M.} and {Giraud-H\'eraud, Y.} and {Gjerl\o{}w, E.} and {Gonz\'alez-Nuevo, J.} and {G\'orski, K. M.} and {Grainge, K. J. B.} and {Gratton, S.} and {Gregorio, A.} and {Gruppuso, A.} and {Gudmundsson, J. E.} and {Hansen, F. K.} and {Hanson, D.} and {Harrison, D. L.} and {Hempel, A.} and {Henrot-Versill\'e, S.} and {Hern\'andez-Monteagudo, C.} and {Herranz, D.} and {Hildebrandt, S. R.} and {Hivon, E.} and {Hobson, M.} and {Holmes, W. A.} and {Hornstrup, A.} and {Hovest, W.} and {Huffenberger, K. M.} and {Hurier, G.} and {Jaffe, A. H.} and {Jaffe, T. R.} and {Jin, T.} and {Jones, W. C.} and {Juvela, M.} and {Keih\"anen, E.} and {Keskitalo, R.} and {Khamitov, I.} and {Kisner, T. S.} and {Kneissl, R.} and {Knoche, J.} and {Kunz, M.} and {Kurki-Suonio, H.} and {Lagache, G.} and {Lamarre, J.-M.} and {Lasenby, A.} and {Lattanzi, M.} and {Lawrence, C. R.} and {Leonardi, R.} and {Lesgourgues, J.} and {Levrier, F.} and {Liguori, M.} and {Lilje, P. B.} and {Linden-V\o{}rnle, M.} and {L\'opez-Caniego, M.} and {Lubin, P. M.} and {Mac\'{\i}as-P\'erez, J. F.} and {Maggio, G.} and {Maino, D.} and {Mak, D. S. Y.} and {Mandolesi, N.} and {Mangilli, A.} and {Martin, P. G.} and {Mart\'{\i}nez-Gonz\'alez, E.} and {Masi, S.} and {Matarrese, S.} and {Mazzotta, P.} and {McGehee, P.} and {Mei, S.} and {Melchiorri, A.} and {Melin, J.-B.} and {Mendes, L.} and {Mennella, A.} and {Migliaccio, M.} and {Mitra, S.} and {Miville-Desch\^enes, M.-A.} and {Moneti, A.} and {Montier, L.} and {Morgante, G.} and {Mortlock, D.} and {Moss, A.} and {Munshi, D.} and {Murphy, J. A.} and {Naselsky, P.} and {Nastasi, A.} and {Nati, F.} and {Natoli, P.} and {Netterfield, C. B.} and {N\o{}rgaard-Nielsen, H. U.} and {Noviello, F.} and {Novikov, D.} and {Novikov, I.} and {Olamaie, M.} and {Oxborrow, C. A.} and {Paci, F.} and {Pagano, L.} and {Pajot, F.} and {Paoletti, D.} and {Pasian, F.} and {Patanchon, G.} and {Pearson, T. J.} and {Perdereau, O.} and {Perotto, L.} and {Perrott, Y. C.} and {Perrotta, F.} and {Pettorino, V.} and {Piacentini, F.} and {Piat, M.} and {Pierpaoli, E.} and {Pietrobon, D.} and {Plaszczynski, S.} and {Pointecouteau, E.} and {Polenta, G.} and {Pratt, G. W.} and {Pr\'ezeau, G.} and {Prunet, S.} and {Puget, J.-L.} and {Rachen, J. P.} and {Reach, W. T.} and {Rebolo, R.} and {Reinecke, M.} and {Remazeilles, M.} and {Renault, C.} and {Renzi, A.} and {Ristorcelli, I.} and {Rocha, G.} and {Rosset, C.} and {Rossetti, M.} and {Roudier, G.} and {Rozo, E.} and {Rubi\~no-Mart\'{\i}n, J. A.} and {Rumsey, C.} and {Rusholme, B.} and {Rykoff, E. S.} and {Sandri, M.} and {Santos, D.} and {Saunders, R. D. E.} and {Savelainen, M.} and {Savini, G.} and {Schammel, M. P.} and {Scott, D.} and {Seiffert, M. D.} and {Shellard, E. P. S.} and {Shimwell, T. W.} and {Spencer, L. D.} and {Stanford, S. A.} and {Stern, D.} and {Stolyarov, V.} and {Stompor, R.} and {Streblyanska, A.} and {Sudiwala, R.} and {Sunyaev, R.} and {Sutton, D.} and {Suur-Uski, A.-S.} and {Sygnet, J.-F.} and {Tauber, J. A.} and {Terenzi, L.} and {Toffolatti, L.} and {Tomasi, M.} and {Tramonte, D.} and {Tristram, M.} and {Tucci, M.} and {Tuovinen, J.} and {Umana, G.} and {Valenziano, L.} and {Valiviita, J.} and {Van Tent, B.} and {Vielva, P.} and {Villa, F.} and {Wade, L. A.} and {Wandelt, B. D.} and {Wehus, I. K.} and {White, S. D. M.} and {Wright, E. L.} and {Yvon, D.} and {Zacchei, A.} and {Zonca, A.}},
	title = {Planck 2015 results - XXVII. The second Planck catalogue of Sunyaev-Zeldovich sources},
	DOI= "10.1051/0004-6361/201525823",
	url= "https://doi.org/10.1051/0004-6361/201525823",
	journal = {A\&A},
	year = 2016,
	volume = 594,
	pages = "A27"}

@article{Voit2005,
       author = {{Voit}, G. Mark},
        title = "{Tracing cosmic evolution with clusters of galaxies}",
      journal = {Reviews of Modern Physics},
         year = 2005,
        month = apr,
       volume = {77},
       number = {1},
        pages = {207-258},
          doi = {10.1103/RevModPhys.77.207},
archivePrefix = {arXiv},
       eprint = {astro-ph/0410173},
 primaryClass = {astro-ph},
       adsurl = {https://ui.adsabs.harvard.edu/abs/2005RvMP...77..207V}}

@ARTICLE{Brunetti2014,
       author = {{Brunetti}, Gianfranco and {Jones}, Thomas W.},
        title = "{Cosmic Rays in Galaxy Clusters and Their Nonthermal Emission}",
      journal = {International Journal of Modern Physics D},
         year = 2014,
        month = mar,
       volume = {23},
       number = {4},
          eid = {1430007-98},
        pages = {1430007-98},
          doi = {10.1142/S0218271814300079},
archivePrefix = {arXiv},
       eprint = {1401.7519},
 primaryClass = {astro-ph.CO},
       adsurl = {https://ui.adsabs.harvard.edu/abs/2014IJMPD..2330007B}}

@BOOK{sarazin1988,
       author = {{Sarazin}, Craig L.},
        title = "{X-ray emission from clusters of galaxies}",
         year = 1988,
    publisher = {American Physical Society},
       adsurl = {https://ui.adsabs.harvard.edu/abs/1988xrec.book.....S}}

@ARTICLE{Markevitch2007,
       author = {{Markevitch}, Maxim and {Vikhlinin}, Alexey},
        title = "{Shocks and cold fronts in galaxy clusters}",
      journal = {\physrep},
         year = 2007,
        month = may,
       volume = {443},
       number = {1},
        pages = {1-53},
          doi = {10.1016/j.physrep.2007.01.001},
archivePrefix = {arXiv},
       eprint = {astro-ph/0701821},
 primaryClass = {astro-ph},
       adsurl = {https://ui.adsabs.harvard.edu/abs/2007PhR...443....1M}}

@ARTICLE{Feretti2012,
       author = {{Feretti}, Luigina and {Giovannini}, Gabriele and {Govoni}, Federica and {Murgia}, Matteo},
        title = "{Clusters of galaxies: observational properties of the diffuse radio emission}",
      journal = {\aapr},
         year = 2012,
        month = may,
       volume = {20},
          eid = {54},
        pages = {54},
          doi = {10.1007/s00159-012-0054-z},
archivePrefix = {arXiv},
       eprint = {1205.1919},
 primaryClass = {astro-ph.CO},
       adsurl = {https://ui.adsabs.harvard.edu/abs/2012A&ARv..20...54F}}

@ARTICLE{Weisskopf2002,
       author = {{Weisskopf}, M.~C. and {Brinkman}, B. and {Canizares}, C. and {Garmire}, G. and {Murray}, S. and {Van Speybroeck}, L.~P.},
        title = "{An Overview of the Performance and Scientific Results from the Chandra X-Ray Observatory}",
      journal = {\pasp},
         year = 2002,
        month = jan,
       volume = {114},
       number = {791},
        pages = {1-24},
          doi = {10.1086/338108},
archivePrefix = {arXiv},
       eprint = {astro-ph/0110308},
 primaryClass = {astro-ph},
       adsurl = {https://ui.adsabs.harvard.edu/abs/2002PASP..114....1W}}

@ARTICLE{Jansen2001,
       author = {{Jansen}, F. and {Lumb}, D. and {Altieri}, B. and {Clavel}, J. and {Ehle}, M. and {Erd}, C. and {Gabriel}, C. and {Guainazzi}, M. and {Gondoin}, P. and {Much}, R. and {Munoz}, R. and {Santos}, M. and {Schartel}, N. and {Texier}, D. and {Vacanti}, G.},
        title = "{XMM-Newton observatory. I. The spacecraft and operations}",
      journal = {\aap},
         year = 2001,
        month = jan,
       volume = {365},
        pages = {L1-L6},
          doi = {10.1051/0004-6361:20000036},
       adsurl = {https://ui.adsabs.harvard.edu/abs/2001A&A...365L...1J}}

@INPROCEEDINGS{Jonas2016,
       author = {{Jonas}, J. and {MeerKAT Team}},
        title = "{The MeerKAT Radio Telescope}",
    booktitle = {MeerKAT Science: On the Pathway to the SKA},
         year = 2016,
        month = jan,
          eid = {1},
        pages = {1},
          doi = {10.22323/1.277.0001},
       adsurl = {https://ui.adsabs.harvard.edu/abs/2016mks..confE...1J}}

@ARTICLE{Pratt2007,
       author = {{Pratt}, G.~W. and {B{\"o}hringer}, H. and {Croston}, J.~H. and {Arnaud}, M. and {Borgani}, S. and {Finoguenov}, A. and {Temple}, R.~F.},
        title = "{Temperature profiles of a representative sample of nearby X-ray galaxy clusters}",
      journal = {\aap},
         year = 2007,
        month = jan,
       volume = {461},
       number = {1},
        pages = {71-80},
          doi = {10.1051/0004-6361:20065676},
archivePrefix = {arXiv},
       eprint = {astro-ph/0609480},
 primaryClass = {astro-ph},
       adsurl = {https://ui.adsabs.harvard.edu/abs/2007A&A...461...71P}}

@article{Ettori2010,
	author = {{Ettori}, Stefano and {Gastaldello}, Fabio and {Leccardi, A.} and {Molendi, S.} and {Rossetti, M.} and {Buote, D.} and {Meneghetti, M.}},
	title = {Mass profiles and c--M$_{\mathrm{DM}}$ relation in X-ray luminous galaxy clusters },
	DOI= "10.1051/0004-6361/201015271",
	url= "https://doi.org/10.1051/0004-6361/201015271",
	journal = {A\&A},
	year = 2010,
	volume = 524,
	pages = "A68"}

@ARTICLE{Kravtsov2006,
       author = {{Kravtsov}, Andrey V. and {Vikhlinin}, Alexey and {Nagai}, Daisuke},
        title = "{A New Robust Low-Scatter X-Ray Mass Indicator for Clusters of Galaxies}",
      journal = {\apj},
         year = 2006,
        month = oct,
       volume = {650},
       number = {1},
        pages = {128-136},
          doi = {10.1086/506319},
archivePrefix = {arXiv},
       eprint = {astro-ph/0603205},
 primaryClass = {astro-ph},
       adsurl = {https://ui.adsabs.harvard.edu/abs/2006ApJ...650..128K}}

@misc{Hugo2021,
  title = {Reference Flux Scale for MeerKAT: Long Term Observation and Field Modelling of PKS B0407-65},
  author = {Hugo, Benjamin V},
  year = {2021}}

@INPROCEEDINGS{McMullin2007,
       author = {{McMullin}, J.~P. and {Waters}, B. and {Schiebel}, D. and {Young}, W. and {Golap}, K.},
        title = "{CASA Architecture and Applications}",
    booktitle = {Astronomical Data Analysis Software and Systems XVI},
         year = 2007,
       editor = {{Shaw}, R.~A. and {Hill}, F. and {Bell}, D.~J.},
       series = {Astronomical Society of the Pacific Conference Series},
       volume = {376},
        month = oct,
        pages = {127},
       adsurl = {https://ui.adsabs.harvard.edu/abs/2007ASPC..376..127M}}

@ARTICLE{CASA2022,
       author = {{CASA Team} and {Bean}, Ben and {Bhatnagar}, Sanjay and {Castro}, Sandra and {Donovan Meyer}, Jennifer and {Emonts}, Bjorn and {Garcia}, Enrique and {Garwood}, Robert and {Golap}, Kumar and {Gonzalez Villalba}, Justo and {Harris}, Pamela and {Hayashi}, Yohei and {Hoskins}, Josh and {Hsieh}, Mingyu and {Jagannathan}, Preshanth and {Kawasaki}, Wataru and {Keimpema}, Aard and {Kettenis}, Mark and {Lopez}, Jorge and {Marvil}, Joshua and {Masters}, Joseph and {McNichols}, Andrew and {Mehringer}, David and {Miel}, Renaud and {Moellenbrock}, George and {Montesino}, Federico and {Nakazato}, Takeshi and {Ott}, Juergen and {Petry}, Dirk and {Pokorny}, Martin and {Raba}, Ryan and {Rau}, Urvashi and {Schiebel}, Darrell and {Schweighart}, Neal and {Sekhar}, Srikrishna and {Shimada}, Kazuhiko and {Small}, Des and {Steeb}, Jan-Willem and {Sugimoto}, Kanako and {Suoranta}, Ville and {Tsutsumi}, Takahiro and {van Bemmel}, Ilse M. and {Verkouter}, Marjolein and {Wells}, Akeem and {Xiong}, Wei and {Szomoru}, Arpad and {Griffith}, Morgan and {Glendenning}, Brian and {Kern}, Jeff},
        title = "{CASA, the Common Astronomy Software Applications for Radio Astronomy}",
      journal = {\pasp},
         year = 2022,
        month = nov,
       volume = {134},
       number = {1041},
          eid = {114501},
        pages = {114501},
          doi = {10.1088/1538-3873/ac9642},
archivePrefix = {arXiv},
       eprint = {2210.02276},
 primaryClass = {astro-ph.IM},
       adsurl = {https://ui.adsabs.harvard.edu/abs/2022PASP..134k4501C}}

@ARTICLE{Botteon2024,
       author = {{Botteon}, A. and {van Weeren}, R.~J. and {Eckert}, D. and {Gastaldello}, F. and {Markevitch}, M. and {Giacintucci}, S. and {Brunetti}, G. and {Kale}, R. and {Venturi}, T.},
        title = "{The prototypical major cluster merger Abell 754: I. Calibration of MeerKAT data and radio/X-ray spectral mapping of the cluster}",
      journal = {\aap},
         year = 2024,
        month = oct,
       volume = {690},
          eid = {A222},
        pages = {A222},
          doi = {10.1051/0004-6361/202451293},
archivePrefix = {arXiv},
       eprint = {2406.18983},
 primaryClass = {astro-ph.CO},
       adsurl = {https://ui.adsabs.harvard.edu/abs/2024A&A...690A.222B}}

@software{Offringa2010,
       author = {{Offringa}, A.~R.},
        title = "{AOFlagger: RFI Software}",
 howpublished = {Astrophysics Source Code Library, record ascl:1010.017},
         year = 2010,
        month = oct,
          eid = {ascl:1010.017},
archivePrefix = {ascl},
       eprint = {1010.017},
       adsurl = {https://ui.adsabs.harvard.edu/abs/2010ascl.soft10017O}}

@article{Offringa2012,
	author = {{Offringa}, A. R. and {van de Gronde}, J. J. and {Roerdink}, J. B. T. M.},
	title = {A morphological algorithm for improving radio-frequency interference detection},
	DOI= "10.1051/0004-6361/201118497",
	url= "https://doi.org/10.1051/0004-6361/201118497",
	journal = {A\&A},
	year = 2012,
	volume = 539,
	pages = "A95",
	month = ""}

@article{Offringa2014,
       author = {{Offringa}, A.~R. and {McKinley}, B. and {Hurley-Walker}, N. and {Briggs}, F.~H. and {Wayth}, R.~B. and {Kaplan}, D.~L. and {Bell}, M.~E. and {Feng}, L. and {Neben}, A.~R. and {Hughes}, J.~D. and {Rhee}, J. and {Murphy}, T. and {Bhat}, N.~D.~R. and {Bernardi}, G. and {Bowman}, J.~D. and {Cappallo}, R.~J. and {Corey}, B.~E. and {Deshpande}, A.~A. and {Emrich}, D. and {Ewall-Wice}, A. and {Gaensler}, B.~M. and {Goeke}, R. and {Greenhill}, L.~J. and {Hazelton}, B.~J. and {Hindson}, L. and {Johnston-Hollitt}, M. and {Jacobs}, D.~C. and {Kasper}, J.~C. and {Kratzenberg}, E. and {Lenc}, E. and {Lonsdale}, C.~J. and {Lynch}, M.~J. and {McWhirter}, S.~R. and {Mitchell}, D.~A. and {Morales}, M.~F. and {Morgan}, E. and {Kudryavtseva}, N. and {Oberoi}, D. and {Ord}, S.~M. and {Pindor}, B. and {Procopio}, P. and {Prabu}, T. and {Riding}, J. and {Roshi}, D.~A. and {Shankar}, N. Udaya and {Srivani}, K.~S. and {Subrahmanyan}, R. and {Tingay}, S.~J. and {Waterson}, M. and {Webster}, R.~L. and {Whitney}, A.~R. and {Williams}, A. and {Williams}, C.~L.},
        title = "{WSCLEAN: an implementation of a fast, generic wide-field imager for radio astronomy}",
      journal = {\mnras},
         year = 2014,
        month = oct,
       volume = {444},
       number = {1},
        pages = {606-619},
          doi = {10.1093/mnras/stu1368},
archivePrefix = {arXiv},
       eprint = {1407.1943},
 primaryClass = {astro-ph.IM},
       adsurl = {https://ui.adsabs.harvard.edu/abs/2014MNRAS.444..606O}}

@ARTICLE{Balboni2025,
       author = {{Balboni}, M. and {Gastaldello}, F. and {Bonafede}, A. and {Botteon}, A. and {Bartalucci}, I. and {Cassano}, R. and {De Grandi}, S. and {Ettori}, S. and {Gaspari}, M. and {Ghizzardi}, S. and et al.},
        title = "{CHEX-MATE: New detections and properties of the radio diffuse emission in massive clusters with MeerKAT}",
      journal = {\aap},
         year = 2026,
        month = mar,
       volume = {707},
          eid = {A143},
        pages = {A143},
          doi = {10.1051/0004-6361/202556148},
archivePrefix = {arXiv},
       eprint = {2507.00133},
 primaryClass = {astro-ph.CO},
       adsurl = {https://ui.adsabs.harvard.edu/abs/2026A&A...707A.143B}}

@INPROCEEDINGS{Briggs1995,
       author = {{Briggs}, D.~S.},
        title = "{High Fidelity Interferometric Imaging: Robust Weighting and NNLS Deconvolution}",
    booktitle = {American Astronomical Society Meeting Abstracts},
         year = 1995,
       series = {American Astronomical Society Meeting Abstracts},
       volume = {187},
        month = dec,
          eid = {112.02},
        pages = {112.02},
       adsurl = {https://ui.adsabs.harvard.edu/abs/1995AAS...18711202B}}

@article{Dalton2006,
author = {Dalton, G. and Caldwell, Martin and Ward, A. and Whalley, Martin and Woodhouse, G and Edeson, R. and Clark, P and Beard, S. and Gallie, Angus and Todd, Stephen and Strachan, J. and Bezawada, Naidu and Sutherland, W. and Emerson, Jim},
year = {2006},
month = {07},
pages = {},
title = {The VISTA infrared camera},
volume = {6269},
journal = {Proceedings of SPIE - The International Society for Optical Engineering},
doi = {10.1117/12.670018}}

@INPROCEEDINGS{Fruscione2006,
       author = {{Fruscione}, Antonella and {McDowell}, Jonathan C. and {Allen}, Glenn E. and {Brickhouse}, Nancy S. and {Burke}, Douglas J. and {Davis}, John E. and {Durham}, Nick and {Elvis}, Martin and {Galle}, Elizabeth C. and {Harris}, Daniel E. and {Huenemoerder}, David P. and {Houck}, John C. and {Ishibashi}, Bish and {Karovska}, Margarita and {Nicastro}, Fabrizio and {Noble}, Michael S. and {Nowak}, Michael A. and {Primini}, Frank A. and {Siemiginowska}, Aneta and {Smith}, Randall K. and {Wise}, Michael},
        title = "{CIAO: Chandra's data analysis system}",
    booktitle = {Observatory Operations: Strategies, Processes, and Systems},
         year = 2006,
       editor = {{Silva}, David R. and {Doxsey}, Rodger E.},
       series = {Society of Photo-Optical Instrumentation Engineers (SPIE) Conference Series},
       volume = {6270},
        month = jun,
          eid = {62701V},
        pages = {62701V},
          doi = {10.1117/12.671760},
       adsurl = {https://ui.adsabs.harvard.edu/abs/2006SPIE.6270E..1VF}}

@ARTICLE{Bourdin2013,
       author = {{Bourdin}, H. and {Mazzotta}, P. and {Markevitch}, M. and {Giacintucci}, S. and {Brunetti}, G.},
        title = "{Shock Heating of the Merging Galaxy Cluster A521}",
      journal = {\apj},
         year = 2013,
        month = feb,
       volume = {764},
       number = {1},
          eid = {82},
        pages = {82},
          doi = {10.1088/0004-637X/764/1/82},
archivePrefix = {arXiv},
       eprint = {1302.0696},
 primaryClass = {astro-ph.CO},
       adsurl = {https://ui.adsabs.harvard.edu/abs/2013ApJ...764...82B}}

@ARTICLE{Bourdin2004,
       author = {{Bourdin}, H. and {Sauvageot}, J. -L. and {Slezak}, E. and {Bijaoui}, A. and {Teyssier}, R.},
        title = "{Temperature map computation for X-ray clusters of galaxies}",
      journal = {\aap},
         year = 2004,
        month = feb,
       volume = {414},
        pages = {429-443},
          doi = {10.1051/0004-6361:20031662},
       adsurl = {https://ui.adsabs.harvard.edu/abs/2004A&A...414..429B}}

@ARTICLE{Bourdin2008,
       author = {{Bourdin}, H. and {Mazzotta}, P.},
        title = "{Temperature structure of the intergalactic medium within seven nearby and bright clusters of galaxies observed with XMM-Newton}",
      journal = {\aap},
         year = 2008,
        month = feb,
       volume = {479},
       number = {2},
        pages = {307-320},
          doi = {10.1051/0004-6361:20065758},
archivePrefix = {arXiv},
       eprint = {0802.1866},
 primaryClass = {astro-ph},
       adsurl = {https://ui.adsabs.harvard.edu/abs/2008A&A...479..307B}}

@ARTICLE{cavaliere1976,
       author = {{Cavaliere}, A. and {Fusco-Femiano}, R.},
        title = "{X-rays from hot plasma in clusters of galaxies.}",
      journal = {\aap},
         year = 1976,
        month = may,
       volume = {49},
        pages = {137-144},
       adsurl = {https://ui.adsabs.harvard.edu/abs/1976A&A....49..137C}}

@ARTICLE{Cavaliere1978,
       author = {{Cavaliere}, A. and {Fusco-Femiano}, R.},
        title = "{The Distribution of Hot Gas in Clusters of Galaxies}",
      journal = {\aap},
         year = 1978,
        month = nov,
       volume = {70},
        pages = {677},
       adsurl = {https://ui.adsabs.harvard.edu/abs/1978A&A....70..677C}
}

@ARTICLE{anderson1989,
       author = {{Anders}, E. and {Grevesse}, N.},
        title = "{Abundances of the elements: Meteoritic and solar}",
      journal = {\gca},
         year = 1989,
        month = jan,
       volume = {53},
       number = {1},
        pages = {197-214},
          doi = {10.1016/0016-7037(89)90286-X},
       adsurl = {https://ui.adsabs.harvard.edu/abs/1989GeCoA..53..197A}}

@article{Lyskova2019,
    author = {Lyskova, N and Churazov, E and Zhang, C and Forman, W and Jones, C and Dolag, K and Roediger, E and Sheardown, A},
    title = {Close-up view of an ongoing merger between the NGC 4839 group and the Coma cluster – a post-merger scenario},
    journal = {Monthly Notices of the Royal Astronomical Society},
    volume = {485},
    number = {2},
    pages = {2922-2934},
    year = {2019},
    month = {03},
    issn = {0035-8711},
    doi = {10.1093/mnras/stz597},
    url = {https://doi.org/10.1093/mnras/stz597},
    eprint = {https://academic.oup.com/mnras/article-pdf/485/2/2922/28075387/stz597.pdf}}

@ARTICLE{ZuHone2018,
       author = {{ZuHone}, J.~A. and {Kowalik}, K. and {{\"O}hman}, E. and {Lau}, E. and {Nagai}, D.},
        title = "{The Galaxy Cluster Merger Catalog: An Online Repository of Mock Observations from Simulated Galaxy Cluster Mergers}",
      journal = {\apjs},
         year = 2018,
        month = jan,
       volume = {234},
       number = {1},
          eid = {4},
        pages = {4},
          doi = {10.3847/1538-4365/aa99db},
archivePrefix = {arXiv},
       eprint = {1609.04121},
 primaryClass = {astro-ph.CO},
       adsurl = {https://ui.adsabs.harvard.edu/abs/2018ApJS..234....4Z}}

@ARTICLE{ZuHone2011,
       author = {{ZuHone}, J.~A.},
        title = "{A Parameter Space Exploration of Galaxy Cluster Mergers. I. Gas Mixing and the Generation of Cluster Entropy}",
      journal = {\apj},
         year = 2011,
        month = feb,
       volume = {728},
       number = {1},
          eid = {54},
        pages = {54},
          doi = {10.1088/0004-637X/728/1/54},
archivePrefix = {arXiv},
       eprint = {1004.3820},
 primaryClass = {astro-ph.CO},
       adsurl = {https://ui.adsabs.harvard.edu/abs/2011ApJ...728...54Z}}

@article{Adam2017_b,
	author = {{Adam}, Rémi and {Arnaud}, Monique and {Bartalucci}, Iacopo and {Ade, P.} and {Andr\'e, P.} and {Beelen, A.} and {Beno\^{\i}t, A.} and {Bideaud, A.} and {Billot, N.} and {Bourdin, H.} and {Bourrion, O.} and {Calvo, M.} and {Catalano, A.} and {Coiffard, G.} and {Comis, B.} and {D\'{}Addabbo, A.} and {D\'esert, F.-X.} and {Doyle, S.} and {Ferrari, C.} and {Goupy, J.} and {Kramer, C.} and {Lagache, G.} and {Leclercq, S.} and {Mac\'{\i}as-P\'erez, J.-F.} and {Maurogordato, S.} and {Mauskopf, P.} and {Mayet, F.} and {Monfardini, A.} and {Pajot, F.} and {Pascale, E.} and {Perotto, L.} and {Pisano, G.} and {Pointecouteau, E.} and {Ponthieu, N.} and {Pratt, G. W.} and {Rev\'eret, V.} and {Ritacco, A.} and {Rodriguez, L.} and {Romero, C.} and {Ruppin, F.} and {Schuster, K.} and {Sievers, A.} and {Triqueneaux, S.} and {Tucker, C.} and {Zylka, R.}},
	title = {Mapping the hot gas temperature in galaxy clusters using X-ray and Sunyaev-Zel\'{}dovich imaging},
	DOI= "10.1051/0004-6361/201629810",
	url= "https://doi.org/10.1051/0004-6361/201629810",
	journal = {A\&A},
	year = 2017,
	volume = 606,
	pages = "A64"}

@article{Bartalucci2024,
	author = {{Bartalucci}, Iacopo and {Rossetti}, Mariachiara and {Boschin}, Walter and {Girardi, M.} and {Nonino, M.} and {Baraldi, E.} and {Balboni, M.} and {Coe, D.} and {De Grandi, S.} and {Gastaldello, F.} and {Ghizzardi, S.} and {Giacintucci, S.} and {Grillo, C.} and {Harvey, D.} and {Lovisari, L.} and {Molendi, S.} and {Resseguier, T.} and {Riva, G.} and {Venturi, T.} and {Zitrin, A.}},
	title = {PSZ2 G282.28+49.94, a recently discovered analogue of the famous Bullet Cluster},
	DOI= "10.1051/0004-6361/202450468",
	url= "https://doi.org/10.1051/0004-6361/202450468",
	journal = {A\&A},
	year = 2024,
	volume = 689,
	pages = "A324"}

@ARTICLE{Eckert2020,
       author = {{Eckert}, Dominique and {Finoguenov}, Alexis and {Ghirardini}, Vittorio and {Grandis}, Sebastian and {Kaefer}, Florian and {Sanders}, Jeremy and {Ramos-Ceja}, Miriam},
        title = "{Low-scatter galaxy cluster mass proxies for the eROSITA all-sky survey}",
      journal = {The Open Journal of Astrophysics},
         year = 2020,
        month = sep,
       volume = {3},
          eid = {12},
        pages = {12},
          doi = {10.21105/astro.2009.13944},
archivePrefix = {arXiv},
       eprint = {2009.03944},
 primaryClass = {astro-ph.CO},
       adsurl = {https://ui.adsabs.harvard.edu/abs/2020OJAp....3E..12E}}

@ARTICLE{Botteon2023,
       author = {{Botteon}, Andrea and {Markevitch}, Maxim and {van Weeren}, Reinout J. and {Brunetti}, Gianfranco and {Shimwell}, Timothy W.},
        title = "{Surface brightness discontinuities in radio halos. Insights from the MeerKAT Galaxy Cluster Legacy Survey}",
      journal = {\aap},
         year = 2023,
        month = jun,
       volume = {674},
          eid = {A53},
        pages = {A53},
          doi = {10.1051/0004-6361/202346150},
archivePrefix = {arXiv},
       eprint = {2302.07881},
 primaryClass = {astro-ph.CO},
       adsurl = {https://ui.adsabs.harvard.edu/abs/2023A&A...674A..53B}
}

@ARTICLE{Cassano2006,
       author = {{Cassano}, R. and {Brunetti}, G. and {Setti}, G.},
        title = "{Statistics of giant radio haloes from electron reacceleration models}",
      journal = {\mnras},
         year = 2006,
        month = jul,
       volume = {369},
       number = {4},
        pages = {1577-1595},
          doi = {10.1111/j.1365-2966.2006.10423.x},
archivePrefix = {arXiv},
       eprint = {astro-ph/0604103},
 primaryClass = {astro-ph},
       adsurl = {https://ui.adsabs.harvard.edu/abs/2006MNRAS.369.1577C}
}

@ARTICLE{Donnert2013,
       author = {{Donnert}, J. and {Dolag}, K. and {Brunetti}, G. and {Cassano}, R.},
        title = "{Rise and fall of radio haloes in simulated merging galaxy clusters}",
      journal = {\mnras},
         year = 2013,
        month = mar,
       volume = {429},
       number = {4},
        pages = {3564-3569},
          doi = {10.1093/mnras/sts628},
archivePrefix = {arXiv},
       eprint = {1211.3337},
 primaryClass = {astro-ph.CO},
       adsurl = {https://ui.adsabs.harvard.edu/abs/2013MNRAS.429.3564D}
}

@ARTICLE{Nishiwaki2026,
	author = {{Nishiwaki, K.} and {Brunetti, G.} and {Vazza, F.} and {Gheller, C.}},
	title = {Cosmological simulation of a radio synchrotron bridge between pre-merging galaxy clusters},
	DOI= "10.1051/0004-6361/202558472",
	url= "https://doi.org/10.1051/0004-6361/202558472",
	journal = {A\&A},
	year = 2026,
	volume = 709,
	pages = "A191",
}

@ARTICLE{Hu2025,
       author = {{Hu}, Dan and {Werner}, Norbert and {Xu}, Haiguang and {Zheng}, Qian and {Breuer}, Jean-Paul and {Wu}, Linhui and {Duchesne}, Stefan W. and {van Weeren}, Reinout J. and {Sun}, Ming and {Zhang}, Congyao and {Johnston-Hollitt}, Melanie and {Shan}, Huanyuan and {Guo}, Quan and {Zhu}, Zhenghao and {Wang}, Jingying and {Gu}, Junhua and {Zhao}, Yuanyuan and {Siew}, Hoongwah and {Mao}, Junjie and {Zhang}, Zhongli and {Pl{\v{s}}ek}, Tom{\'a}{\v{s}}},
        title = "{MeerKAT discovery of gigahertz radio emission extending from Abell 3017 towards Abell 3016}",
      journal = {\aap},
         year = 2025,
        month = feb,
       volume = {694},
          eid = {A320},
        pages = {A320},
          doi = {10.1051/0004-6361/202453200},
archivePrefix = {arXiv},
       eprint = {2412.00204},
 primaryClass = {astro-ph.CO},
       adsurl = {https://ui.adsabs.harvard.edu/abs/2025A&A...694A.320H}
}

@ARTICLE{Pignataro2024,
       author = {{Pignataro}, G.~V. and {Bonafede}, A. and {Bernardi}, G. and {Balboni}, M. and {Vazza}, F. and {van Weeren}, R.~J. and {Ubertosi}, F. and {Cassano}, R. and {Brunetti}, G. and {Botteon}, A. and {Venturi}, T. and {Akamatsu}, H. and {Drabent}, A. and {Hoeft}, M.},
        title = "{Mind the gap between A2061 and A2067: Unveiling new diffuse, large-scale radio emission}",
      journal = {\aap},
         year = 2024,
        month = nov,
       volume = {691},
          eid = {A99},
        pages = {A99},
          doi = {10.1051/0004-6361/202451529},
archivePrefix = {arXiv},
       eprint = {2409.15412},
 primaryClass = {astro-ph.CO},
       adsurl = {https://ui.adsabs.harvard.edu/abs/2024A&A...691A..99P}
}

@ARTICLE{Sunyaev1972,
       author = {{Sunyaev}, R.~A. and {Zeldovich}, Ya. B.},
        title = "{The Observations of Relic Radiation as a Test of the Nature of X-Ray Radiation from the Clusters of Galaxies}",
      journal = {Comments on Astrophysics and Space Physics},
         year = 1972,
        month = nov,
       volume = {4},
        pages = {173},
       adsurl = {https://ui.adsabs.harvard.edu/abs/1972CoASP...4..173S}}

@ARTICLE{Ruppin2020,
       author = {{Ruppin}, F. and {McDonald}, M. and {Brodwin}, M. and {Adam}, R. and {Ade}, P. and {Andr{\'e}}, P. and {Andrianasolo}, A. and {Arnaud}, M. and {Aussel}, H. and {Bartalucci}, I. and {Bautz}, M.~W. and {Beelen}, A. and {Beno{\^\i}t}, A. and {Bideaud}, A. and {Bourrion}, O. and {Calvo}, M. and {Catalano}, A. and {Comis}, B. and {Decker}, B. and {De Petris}, M. and {D{\'e}sert}, F.-X. and {Doyle}, S. and {Driessen}, E.~F.~C. and {Eisenhardt}, P.~R.~M. and {Gomez}, A. and {Gonzalez}, A.~H. and {Goupy}, J. and {K{\'e}ruzor{\'e}}, F. and {Kramer}, C. and {Ladjelate}, B. and {Lagache}, G. and {Leclercq}, S. and {Lestrade}, J.-F. and {Mac{\'\i}as-P{\'e}rez}, J.~F. and {Mauskopf}, P. and {Mayet}, F. and {Monfardini}, A. and {Moravec}, E. and {Perotto}, L. and {Pisano}, G. and {Pointecouteau}, E. and {Ponthieu}, N. and {Pratt}, G.~W. and {Rev{\'e}ret}, V. and {Ritacco}, A. and {Romero}, C. and {Roussel}, H. and {Schuster}, K. and {Shu}, S. and {Sievers}, A. and {Stanford}, S.~A. and {Stern}, D. and {Tucker}, C. and {Zylka}, R.},
        title = "{Unveiling the Merger Dynamics of the Most Massive MaDCoWS Cluster at z = 1.2 from a Multiwavelength Mapping of Its Intracluster Medium Properties}",
      journal = {\apj},
         year = 2020,
        month = apr,
       volume = {893},
       number = {1},
          eid = {74},
        pages = {74},
          doi = {10.3847/1538-4357/ab8007},
archivePrefix = {arXiv},
       eprint = {1911.00560},
 primaryClass = {astro-ph.CO},
       adsurl = {https://ui.adsabs.harvard.edu/abs/2020ApJ...893...74R}}

@ARTICLE{Girardi2005,
       author = {{Girardi}, M. and {Demarco}, R. and {Rosati}, P. and {Borgani}, S.},
        title = "{Internal dynamics of the z \raisebox{-0.5ex}\textasciitilde 0.8 cluster RX J0152.7-1357}",
      journal = {\aap},
         year = 2005,
        month = oct,
       volume = {442},
       number = {1},
        pages = {29-41},
          doi = {10.1051/0004-6361:20053232},
archivePrefix = {arXiv},
       eprint = {astro-ph/0506211},
 primaryClass = {astro-ph},
       adsurl = {https://ui.adsabs.harvard.edu/abs/2005A&A...442...29G}}

@ARTICLE{Pandge2019,
       author = {{Pandge}, M.~B. and {Monteiro-Oliveira}, R. and {Bagchi}, J. and {Simionescu}, A. and {Limousin}, M. and {Raychaudhury}, S.},
        title = "{A combined X-ray, optical, and radio view of the merging galaxy cluster MACS J0417.5-1154}",
      journal = {\mnras},
         year = 2019,
        month = feb,
       volume = {482},
       number = {4},
        pages = {5093-5105},
          doi = {10.1093/mnras/sty2937},
archivePrefix = {arXiv},
       eprint = {1810.12071},
 primaryClass = {astro-ph.CO},
       adsurl = {https://ui.adsabs.harvard.edu/abs/2019MNRAS.482.5093P}}

@ARTICLE{Lupton2004,
       author = {{Lupton}, Robert and {Blanton}, Michael R. and {Fekete}, George and {Hogg}, David W. and {O'Mullane}, Wil and {Szalay}, Alex and {Wherry}, Nicholas},
        title = "{Preparing Red-Green-Blue Images from CCD Data}",
      journal = {\pasp},
         year = 2004,
        month = feb,
       volume = {116},
       number = {816},
        pages = {133-137},
          doi = {10.1086/382245},
archivePrefix = {arXiv},
       eprint = {astro-ph/0312483},
 primaryClass = {astro-ph},
       adsurl = {https://ui.adsabs.harvard.edu/abs/2004PASP..116..133L}}

@ARTICLE{Hou2009,
       author = {{Hou}, Annie and {Parker}, Laura C. and {Harris}, William E. and {Wilman}, David J.},
        title = "{Statistical Tools for Classifying Galaxy Group Dynamics}",
      journal = {\apj},
         year = 2009,
        month = sep,
       volume = {702},
       number = {2},
        pages = {1199-1210},
          doi = {10.1088/0004-637X/702/2/1199},
archivePrefix = {arXiv},
       eprint = {0908.0938},
 primaryClass = {astro-ph.GA},
       adsurl = {https://ui.adsabs.harvard.edu/abs/2009ApJ...702.1199H}}

@article{Kass1995,
  title={Bayes factors},
  author={{Kass}, R. and {Raftery}, A.},
  journal={Journal of the american statistical association},
  volume={90},
  number={430},
  pages={773--795},
  year={1995},
  publisher={Taylor \& Francis}}

@article{Scrucca2016,
author = {{Scrucca}, L. and {Fop}, M. and {Murphy}, T. and {Raftery}, A.},
year = {2016},
month = {08},
pages = {205-233},
title = {mclust 5: Clustering, Classiﬁcation and Density Estimation Using Gaussian Finite Mixture Models},
volume = {8},
journal = {The R Journal},
doi = {10.32614/RJ-2016-021}}

@ARTICLE{DS1988,
       author = {{Dressler}, Alan and {Shectman}, Stephen A.},
        title = "{Evidence for Substructure in Rich Clusters of Galaxies from Radial-Velocity Measurements}",
      journal = {\aj},
         year = 1988,
        month = apr,
       volume = {95},
        pages = {985},
          doi = {10.1086/114694},
       adsurl = {https://ui.adsabs.harvard.edu/abs/1988AJ.....95..985D}}

@ARTICLE{Pinkney1996,
       author = {{Pinkney}, Jason and {Roettiger}, Kurt and {Burns}, Jack O. and {Bird}, Christina M.},
        title = "{Evaluation of Statistical Tests for Substructure in Clusters of Galaxies}",
      journal = {\apjs},
         year = 1996,
        month = may,
       volume = {104},
        pages = {1},
          doi = {10.1086/192290},
       adsurl = {https://ui.adsabs.harvard.edu/abs/1996ApJS..104....1P}}

@ARTICLE{Gastaldello2013,
       author = {{Gastaldello}, Fabio and {Di Gesu}, Laura and {Ghizzardi}, Simona and {Giacintucci}, Simona and {Girardi}, Marisa and {Roediger}, Elke and {Rossetti}, Mariachiara and {Brighenti}, Fabrizio and {Buote}, David A. and {Eckert}, Dominique and {Ettori}, Stefano and {Humphrey}, Philip J. and {Mathews}, William G.},
        title = "{Sloshing Cold Fronts in Galaxy Groups and their Perturbing Disk Galaxies: An X-Ray, Optical, and Radio Case Study}",
      journal = {\apj},
         year = 2013,
        month = jun,
       volume = {770},
       number = {1},
          eid = {56},
        pages = {56},
          doi = {10.1088/0004-637X/770/1/56},
archivePrefix = {arXiv},
       eprint = {1304.5478},
 primaryClass = {astro-ph.CO},
       adsurl = {https://ui.adsabs.harvard.edu/abs/2013ApJ...770...56G}}

@ARTICLE{Benavides2023,
       author = {{Benavides}, Jos{\'e} A. and {Biviano}, Andrea and {Abadi}, Mario G.},
        title = "{DS+: A method for the identification of cluster substructures}",
      journal = {\aap},
         year = 2023,
        month = jan,
       volume = {669},
          eid = {A147},
        pages = {A147},
          doi = {10.1051/0004-6361/202245422},
archivePrefix = {arXiv},
       eprint = {2212.00040},
 primaryClass = {astro-ph.GA},
       adsurl = {https://ui.adsabs.harvard.edu/abs/2023A&A...669A.147B}}

@ARTICLE{Golovich2016,
       author = {{Golovich}, Nathan and {Dawson}, William A. and {Wittman}, David and {Ogrean}, Georgiana and {van Weeren}, Reinout and {Bonafede}, Annalisa},
        title = "{MC$^{2}$: Dynamical Analysis of the Merging Galaxy Cluster MACS J1149.5+2223}",
      journal = {\apj},
         year = 2016,
        month = nov,
       volume = {831},
       number = {1},
          eid = {110},
        pages = {110},
          doi = {10.3847/0004-637X/831/1/110},
archivePrefix = {arXiv},
       eprint = {1608.01329},
 primaryClass = {astro-ph.CO},
       adsurl = {https://ui.adsabs.harvard.edu/abs/2016ApJ...831..110G}}

@ARTICLE{Golovich2019,
       author = {{Golovich}, N. and {Dawson}, W.~A. and {Wittman}, D.~M. and {Jee}, M.~J. and {Benson}, B. and {Lemaux}, B.~C. and {van Weeren}, R.~J. and {Andrade-Santos}, F. and {Sobral}, D. and {de Gasperin}, F. and {Br{\"u}ggen}, M. and {Brada{\v{c}}}, M. and {Finner}, K. and {Peter}, A.},
        title = "{Merging Cluster Collaboration: Optical and Spectroscopic Survey of a Radio-selected Sample of 29 Merging Galaxy Clusters}",
      journal = {\apjs},
         year = 2019,
        month = feb,
       volume = {240},
       number = {2},
          eid = {39},
        pages = {39},
          doi = {10.3847/1538-4365/aaf88b},
       adsurl = {https://ui.adsabs.harvard.edu/abs/2019ApJS..240...39G}}

@ARTICLE{Botteon2019,
       author = {{Botteon}, A. and {Cassano}, R. and {Eckert}, D. and {Brunetti}, G. and {Dallacasa}, D. and {Shimwell}, T.~W. and {van Weeren}, R.~J. and {Gastaldello}, F. and {Bonafede}, A. and {Br{\"u}ggen}, M. and {B{\^\i}rzan}, L. and {Clavico}, S. and {Cuciti}, V. and {de Gasperin}, F. and {De Grandi}, S. and {Ettori}, S. and {Ghizzardi}, S. and {Rossetti}, M. and {R{\"o}ttgering}, H.~J.~A. and {Sereno}, M.},
        title = "{Particle acceleration in a nearby galaxy cluster pair: the role of cluster dynamics}",
      journal = {\aap},
         year = 2019,
        month = oct,
       volume = {630},
          eid = {A77},
        pages = {A77},
          doi = {10.1051/0004-6361/201936022},
archivePrefix = {arXiv},
       eprint = {1908.07527},
 primaryClass = {astro-ph.CO},
       adsurl = {https://ui.adsabs.harvard.edu/abs/2019A&A...630A..77B}}

@ARTICLE{Clavico2019,
       author = {{Clavico}, S. and {De Grandi}, S. and {Ghizzardi}, S. and {Rossetti}, M. and {Molendi}, S. and {Gastaldello}, F. and {Girardi}, M. and {Boschin}, W. and {Botteon}, A. and {Cassano}, R. and {Br{\"u}ggen}, M. and {Brunetti}, G. and {Dallacasa}, D. and {Eckert}, D. and {Ettori}, S. and {Gaspari}, M. and {Sereno}, M. and {Shimwell}, T. and {van Weeren}, R.~J.},
        title = "{Growth and disruption in the Lyra complex}",
      journal = {\aap},
         year = 2019,
        month = dec,
       volume = {632},
          eid = {A27},
        pages = {A27},
          doi = {10.1051/0004-6361/201936467},
archivePrefix = {arXiv},
       eprint = {1908.02276},
 primaryClass = {astro-ph.CO},
       adsurl = {https://ui.adsabs.harvard.edu/abs/2019A&A...632A..27C}}

@ARTICLE{Einasto2012,
       author = {{Einasto}, M. and {Vennik}, J. and {Nurmi}, P. and {Tempel}, E. and {Ahvensalmi}, A. and {Tago}, E. and {Liivam{\"a}gi}, L.~J. and {Saar}, E. and {Hein{\"a}m{\"a}ki}, P. and {Einasto}, J. and {Mart{\'\i}nez}, V.~J.},
        title = "{Multimodality in galaxy clusters from SDSS DR8: substructure and velocity distribution}",
      journal = {\aap},
         year = 2012,
        month = apr,
       volume = {540},
          eid = {A123},
        pages = {A123},
          doi = {10.1051/0004-6361/201118697},
archivePrefix = {arXiv},
       eprint = {1202.4927},
 primaryClass = {astro-ph.CO},
       adsurl = {https://ui.adsabs.harvard.edu/abs/2012A&A...540A.123E}}

\end{document}